%% file: main.tex
\documentclass[aps,prl,reprint,superscriptaddress]{revtex4-2}

\usepackage{color}
\usepackage{graphicx}
\usepackage{siunitx}

\begin{document}

\title{Ising-like critical behavior of vortex lattices in an active fluid}

\author{Henning Reinken}
\email{henning.reinken@itp.tu-berlin.de}
\affiliation{Technische Universit\"at Berlin, Institute of Theoretical Physics, Stra{\ss}e des 17. Juni 135, 10623 Berlin, Germany} 

\author{Sebastian Heidenreich}
\affiliation{Department of Mathematical Modelling and Data Analysis,
Physikalisch-Technische Bundesanstalt Braunschweig und Berlin, Abbestr. 2-12,
10587 Berlin, Germany}

\author{Markus B\"ar}
\affiliation{Department of Mathematical Modelling and Data Analysis,
Physikalisch-Technische Bundesanstalt Braunschweig und Berlin, Abbestr. 2-12,
10587 Berlin, Germany}

\author{Sabine H. L. Klapp}
\affiliation{Technische Universit\"at Berlin, Institute of Theoretical Physics, Stra{\ss}e des 17. Juni 135, 10623 Berlin, Germany} 

\date{\today}

\begin{abstract}
Turbulent vortex structures emerging in bacterial active fluids can be organized into regular vortex lattices by weak geometrical constraints such as obstacles.
Here we show, using a continuum-theoretical approach, that the formation and destruction of these patterns exhibit features of a continuous second-order equilibrium phase transition,
including long-range correlations, divergent susceptibility, and critical slowing down. The emerging vorticity field can be mapped onto a two-dimensional (2D) Ising model with antiferromagnetic nearest-neighbor interactions by coarse-graining.
The resulting effective temperature is found to be proportional to the strength of the nonlinear advection in the continuum model.
\end{abstract}


\maketitle

\paragraph*{Introduction.}

The nature of transitions between qualitatively different collective states of non-equilibrium matter continues to be an open question in many areas of physics and related fields.
In classical pattern formation and fluid dynamics, the transition from regular waves and patterns to irregular chaotic or turbulent states has been explored in many different areas ranging from pattern formation in reaction-diffusion and fluid systems to the onset of turbulent flows. 
In many of these systems the replacement of periodic patterns was found to be linked to instabilities of the regular structures.
Prominent examples are different types of waves in extended oscillatory systems~\cite{aranson2002world} or spirals~\cite{bar2004breakup,wheeler2006computation} and Turing patterns ~\cite{halatek2018rethinking} in reaction-diffusion systems. 
For the classical problem of the transition from laminar to turbulent pipe flow this approach cannot be applied, since the laminar flow is stable for arbitrarily large Reynolds numbers. Recent large-scale experimental and numerical studies in pipe flows~\cite{avila2011onset,barkley2015rise,barkley2016theoretical} have instead revealed that the transition bears analogies with a non-equilibrium phase transition in the directed percolation class~\cite{hinrichsen2000non}, where the laminar flow corresponds to an absorbing state. 
Transitions to turbulence in other macroscopic flow systems such as Couette flow~\cite{lemoult2016directed}, channel flow~\cite{sano2016universal} or turbulent liquid crystals~\cite{takeuchi2007directed} are exhibiting similar features.
In addition, theoretical studies on other non-equilibrium systems such as coupled chaotic maps have shown that equilibrium-like transitions can occur on larger scales even if the underlying dynamics is deterministic and highly irreversible \cite{miller1993macroscopic,egolf1998dynamical,marcq1997universality,egolf2000equilibrium}.

Here, we focus on non-equilibrium transitions appearing in active matter, the latter representing a new central field of physics showing intriguing forms of collective motion~\cite{marchetti2013hydrodynamics,cates2015motility,bechinger2016active, doostmohammadi2018active,levis2019activity,liao2020dynamical,chate2020dry,bar2020self,gompper20202020}.
In particular, active matter can also display turbulent-like behavior~\cite{alert2021active}.
Remarkably, a recent simulation study~\cite{doostmohammadi2017onset} of the onset of turbulence in an active nematic has yielded striking analogies to directed percolation.
In the present work, we consider, as a key example, the transition between regular vortex patterns~\cite{wioland2013confinement,lushi14fluid,wioland2016ferromagnetic}
and mesoscale turbulence~\cite{wensink2012meso,bratanov2015new} that has been found experimentally in active (bacterial) suspensions such as \textit{Bacillus Subtilis} and colloidal (e.g., Janus-particle) systems~\cite{nishiguchi2015mesoscopic}.
Mesoscale (or low-Reynolds number) turbulence implies a highly dynamical flow field with spiral-like structures, i.e., vortices, that are characterized by a preferred length scale~\cite{heidenreich2016hydrodynamic}.
Exposing such a suspension to geometrical confinement \cite{theillard2017geometric,zhang2020oscillatory,beppu2017geometry}, one may observe regular vortex lattices where both, the vortex centers and their direction of rotation is ordered. 
A striking example occurs in 2D systems of connected chambers \cite{wioland2013confinement,lushi14fluid,wioland2016ferromagnetic}
where neighboring vortices exhibit the same (ferromagnetic) or the opposite spinning direction (antiferromagnetic), resembling a non-equilibrium magnetic spin lattice.
Intriguingly, however, ordered vortex patterns can also emerge in the presence of small obstacles~\cite{nishiguchi2018engineering,sone2019anomalous,reinken2020organizing,zhang2020oscillatory}
and even spontaneously~\cite{james2018turbulence,james2021emergence,reinken2019anisotropic}.
The question then is: How does the vortex pattern arise (or dissolve) from the turbulent state?

In this work we apply a continuum-theoretical approach to investigate an active suspension subject to a square lattice of obstacles with a lattice constant
comparable to the intrinsic length scale in the unconfined system.
Previous research has shown that such an ``external field'' can stabilize antiferromagnetic vortex patterns under conditions where the unconstrained system exhibits turbulence~\cite{nishiguchi2018engineering}, with quantitative agreement between continuum theory and experiment~\cite{reinken2020organizing}. 
A key parameter is the strength of nonlinear advection, $\lambda$, that crucially depends on the stresses generated by the swimming force and the self-propulsion speed~\cite{heidenreich2016hydrodynamic,reinken2018derivation}, which can be tuned, e.g., by changing the oxygen concentration in experiments~\cite{sokolov2012physical}.
To explore the nature of the transition in the obstacle system, we describe the ordered state as a magnetic spin lattice.
We explicitly include the possibility of a disordered (turbulent) state, going beyond earlier work using the spin picture. 
Because the vortices in our system are spatially ``pinned'', the present transition is different from that 
considered in \cite{james2018turbulence,james2021emergence} which focuses on the melting of a spontaneously formed vortex crystal appearing at extreme $\lambda$.
Our results provide strong evidence that the non-equilibrium order-disorder transition can be described as a second-order transition with critical exponents consistent with the 2D Ising universality class, and $\lambda$ playing the role of an effective temperature.

\begin{figure*}
\centering
\includegraphics[width=0.99\linewidth]{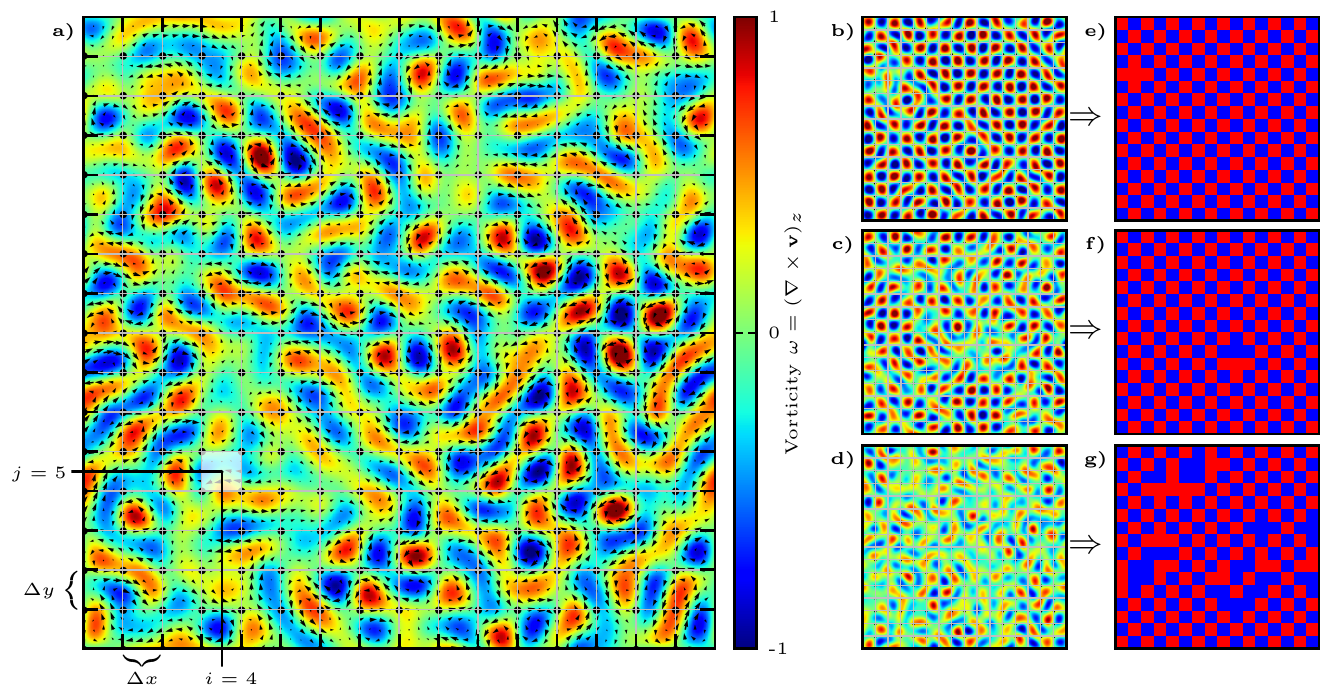}
\caption{\label{fig: snapshots} Snapshots of the vorticity field $\omega$ in a system of size $N = 16\times 16$ obstacles at \textbf{a)} $\lambda = 9.4$ (near the critical point), \textbf{b)} $\lambda = 6$, \textbf{c)} $\lambda = 8$ and at \textbf{d)} $\lambda = 10$. The arrows in \textbf{a)} denote the velocity field $\mathbf{v}$ and the black circles denote the locations of the obstacles. Gray lines indicate the grid used to calculate spin values. The instantaneous spin lattices shown in \textbf{e)}, \textbf{f)} and \textbf{g)} (red for positive, blue for negative spins) are obtained from the vorticity fields in \textbf{b)}, \textbf{c)} and \textbf{d)}.}
\end{figure*}

\paragraph*{Model.} 
We use a well-established minimal model for dense microswimmer suspensions~\cite{dunkel2013fluid,dunkel2013minimal,reinken2018derivation,reinken2019anisotropic,james2018turbulence}, where density fluctuations can be neglected~\cite{be2020phase,zantop2021multi,qi2021active} and the dynamics is described on a coarse-grained (order parameter) level via an effective microswimmer velocity field $\mathbf{v}$~\cite{reinken2018derivation}.
Due to this effective description and the quasi-2D system (where boundaries can act as momentum sinks), momentum is not conserved.
We choose this model over different models with momentum conservation or varying density that have been shown to exhibit similar pattern formation~\cite{grossmann2014vortex,slomka2017geometry,sone2019anomalous}, because it can be derived from microscopic dynamics~\cite{reinken2018derivation} and has been shown to capture experiments on bacteria in the absence~\cite{wensink2012meso} as well as in the presence of obstacle lattices~\cite{reinken2020organizing}.
The dynamics of $\mathbf{v}$ is given by
\begin{eqnarray}
\label{eq: dynamic equation}
&\displaystyle\partial_t \mathbf{v} + \lambda \mathbf{v}\cdot\nabla\mathbf{v} = -\frac{\delta \mathcal{F}}{\delta \mathbf{v}},  \\
&\displaystyle \mathcal{F} =  \int d\mathbf{x} \left[q \nabla\cdot\mathbf{v} - \frac{a}{2} |\mathbf{v}|^2 + \frac{b}{4} |\mathbf{v}|^4  +  \frac{1}{2}\left|(1 + \nabla^2)\mathbf{v}\right|^2 \right]. \nonumber
\end{eqnarray}
The dynamics is characterized by the competition between nonlinear advection ($\lambda \mathbf{v}\cdot\nabla\mathbf{v}$) and relaxation governed by the functional $\mathcal{F}$.
For sufficiently high activity, $0<a<1$, the minimum of $\mathcal{F}$ is a vortex pattern with square lattice symmetry characterized by two perpendicular modes with characteristic wavelength $\Lambda = 2 \pi$ (see~\cite{reinken2019anisotropic,james2018turbulence} and Supplemental Material (SM) for details, which includes Refs.~\cite{odor2004universality,bruce1985universality,bialas2000percolation,hohenberg1977theory,wolgemuth2008collective,bertin2009hydrodynamic,martins2007universality,kenna2008scaling,grinstein1984lower}).
For $\lambda > \lambda^\star$, the nonlinear advection term destabilizes this non-fluctuating ``ground state'' and induces a dynamical state denoted as mesoscale turbulence~\cite{wensink2012meso,dunkel2013fluid,heidenreich2016hydrodynamic,reinken2018derivation,james2018vortex,james2018turbulence}.
Strikingly, recent experiments~\cite{nishiguchi2018engineering} and numerical calculations~\cite{reinken2020organizing} have consistently shown that periodic arrangements of small obstacles can stabilize regular patterns for intermediate values of $\lambda \gtrsim \lambda^\star$.

\paragraph*{Setup.} 
To investigate the transition to a disordered state, i.e., the breakdown of global order for larger values of $\lambda$, we analyze the dynamics in a 2D periodic system containing $N = n_x\times n_y$ obstacles of diameter $\ell=0.13\Lambda$ arranged in a square pattern with lattice constant $L=\sqrt{2}\Lambda$, see Fig.~\ref{fig: snapshots}.
This conforms with the ``ground state'' symmetries and corresponds to the optimal spacing in the sense that it fits to its characteristic scale $\Lambda$~\cite{reinken2020organizing}.
The results are robust with respect to changes of the setup that preserve the general symmetry, e.g., changing $\ell$ or $L$ or adding defects by removing a few obstacles randomly, see SM.
Our study here is confined to the case of square vortex lattices since they have been found to be stable for the case of vanishing nonlinear advection, $\lambda = 0$, whereas vortex lattices with different symmetry, e.g., triangular arrangements~\cite{reinken2020organizing}, are unstable already for $\lambda = 0$.
Hence, the square vortex lattice is the only known example that represents a true minimum of the functional $\mathcal{F}$ and as such may be interpreted as the ``ground state'' of the system with $\lambda<\lambda^\star$. 
We numerically solve Eq.~(\ref{eq: dynamic equation}) using a pseudo-spectral method and implement the obstacles via a local damping potential for both, velocity $\mathbf{v}$ and vorticity $\omega = (\nabla \times \mathbf{v})_z$ (see SM and~\cite{reinken2020organizing} for more details). 
We use system sizes of $N=16\times 16$, $N=32\times 32$ and $N=64\times 64$ and set $a=0.5$, $b=1.6$ as in~\cite{reinken2020organizing}.
To characterize the vortex patterns, we divide the system into a quadratic grid with obstacles occupying the nodes and grid cells numbered by their position in $x$- and $y$-direction, i.e., $i = 1 \dots n_x$ and $j = 1 \dots n_y$, respectively.
We calculate the mean vorticity $\Omega_{i,j}$ in every grid cell via integration of the vorticity field $\omega(x,y) = \big[\nabla \times \mathbf{v}(x,y)\big]_z$,
\begin{equation}
\label{eq: vorticity in cells}
\Omega_{i,j} = \frac{1}{\Delta x \Delta y}\int_{(i-0.5)\Delta x}^{(i+0.5)\Delta x} \int_{(j-0.5)\Delta y}^{(j+0.5)\Delta y} \omega(x,y) \,d x \,d y\, ,
\end{equation}
where $\Delta x$ and $\Delta y$ are the dimensions of the cells.
Normalization yields a system of discrete spins $S_{i,j} = \Omega_{i,j}/|\Omega_{i,j}|$ with $S_{i,j} =\pm 1$.
Using continuous spins $\Omega_{i,j}$ (thus retaining the magnitude of the vorticity) instead of $S_{i,j}$ does not change the nature of the transition, see SM.

\begin{figure}
\includegraphics[width=0.99\linewidth]{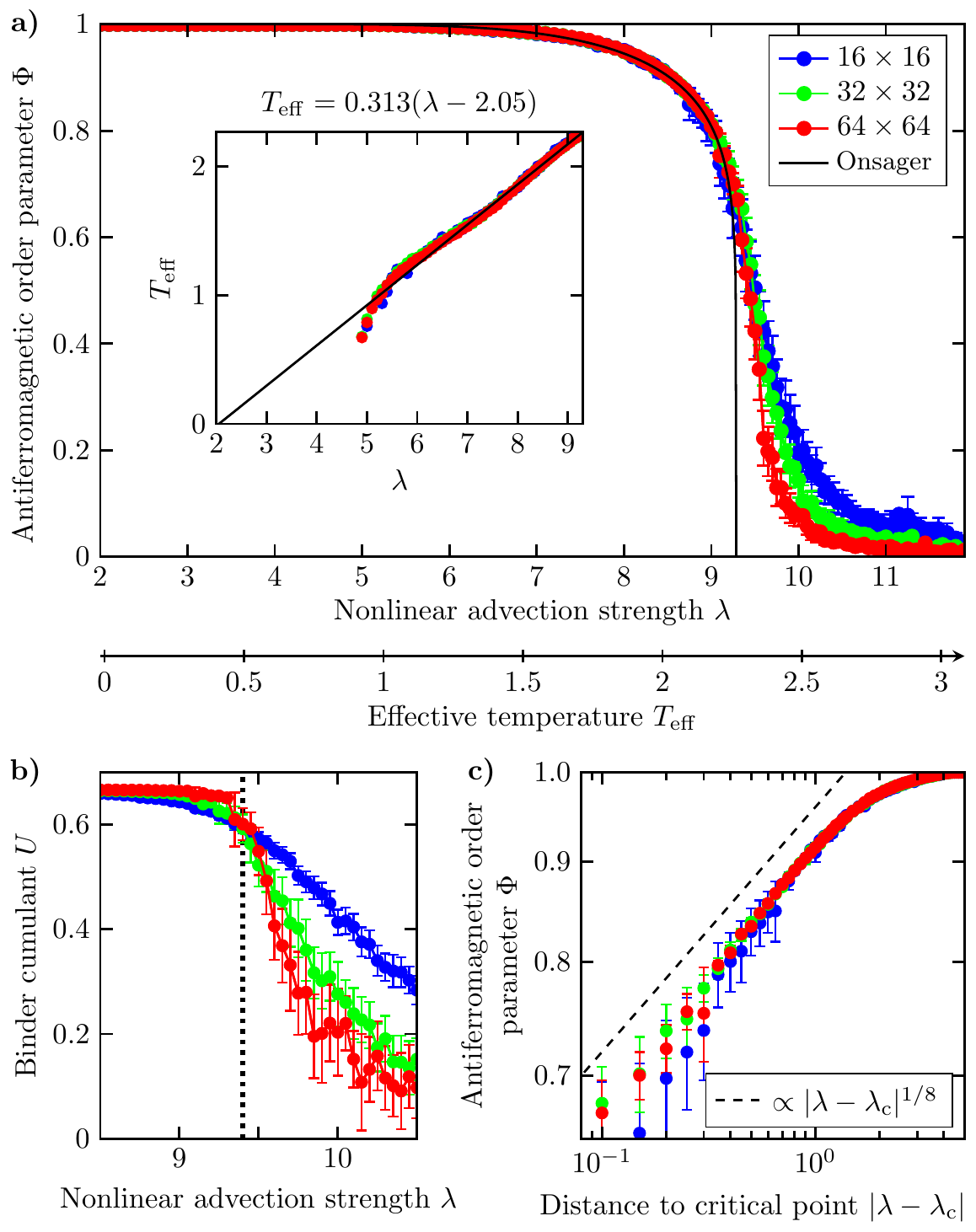}
\caption{\label{fig: order}\textbf{a)} Antiferromagnetic order parameter $\Phi$ as a function of $\lambda$ for different $N$. Solid black line: Onsager's analytical solution as a function of effective temperature $T_\mathrm{eff}(\lambda)$ (shown as second x-axis). Inset: $T_\mathrm{eff}(\lambda)$. \textbf{b)} Binder cumulant $U$ as a function of $\lambda$. The intersection point of the curves for different $N$ marks the critical point $\lambda_\mathrm{c} \approx 9.4$, here denoted by the dotted line. \textbf{c)} $\Phi$ as a function of $|\lambda - \lambda_\mathrm{c}|$. Power-law behavior with exponent $\beta = 1/8$ is shown as a dashed line. The error bars represent the standard error.}
\end{figure}

\paragraph*{Antiferromagnetic order.} 
In analogy to antiferromagnetic spin models, we divide the system into two sublattices (denoted by ${+}$ and ${-}$) and calculate averages for these sublattices separately.
The spatial average of the sublattice spins yields a quantity analogous to a sublattice magnetization per lattice site $m_\pm$, i.e, 
\begin{equation}
\label{eq: magnetization}
m_\pm = \frac{1}{N}\sum_{i=1}^{n_x} \sum_{j=1}^{n_y} \left(1\pm(-1)^{(i+j)}\right) S_{i,j} \, .
\end{equation}
The degree of antiferromagnetic order is measured by the order parameter $\Phi = | \langle m_{+} \rangle - \langle m_{-}\rangle | /2 $, where $\langle \dots \rangle$ denotes the temporal average.
In Fig.~\ref{fig: order}a), $\Phi$ is plotted as a function of nonlinear advection strength $\lambda$.
The data indicate a continuous transition from antiferromagnetic order ($\Phi=1$) at lower values of $\lambda$ to disorder ($\Phi\approx 0$) upon increase of $\lambda$, see also Figs.~\ref{fig: snapshots}b) - d).
The overall behavior is reminiscent of a second-order phase transition.
To locate the critical point occurring at some critical value $\lambda_\mathrm{c}$, we calculate the Binder cumulants, defined via $U_\pm = 1 - \langle m_\pm^4 \rangle/(3\langle m_\pm^2 \rangle^2)$ for the two sublattices, respectively~\cite{binder1981finite}.
Fig.~\ref{fig: order}b) shows the average over the sublattices, $U = (U_{+} + U_{-})/2$, as a function of $\lambda$ for lattices of different size $N$.
At the critical point, the Binder cumulant related to an equilibrium system is known to become independent of $N$, i.e., the curves intersect~\cite{binder1981finite}.
The clear intersection point visible in Fig.~\ref{fig: order}b) implies that we can utilize this method in the present case as well, yielding $\lambda_\mathrm{c} \approx 9.4$.
Moreover, when plotting $\Phi$ as a function of the distance to the critical point $|\lambda - \lambda_\mathrm{c}|$, see Fig.~\ref{fig: order}c), we observe power-law behavior, i.e., $\Phi \propto |\lambda - \lambda_\mathrm{c}|^\beta$, with an exponent $\beta = 1/8$. This conforms with Onsager's exact solution for the magnetization of the 2D Ising model with nearest-neighbor interactions~\cite{onsager1944two},
\begin{equation}
\label{eq: Onsager's solution}
\langle m \rangle = \left\{1 - \sinh^{-4}\left[2J(k_\mathrm{B}T)^{-1}\right] \right\}^{\frac{1}{8}}\, ,
\end{equation}
with $(k_\mathrm{B}T)^{-1}$ and $J$ representing the inverse thermal energy and interaction strength, respectively.
Due to the bi-partite nature of the square lattice, ferro- or antiferromagnetic interactions lead to the same general behavior (in the absence of an external field); $\langle m \rangle$ thus corresponds to our antiferromagnetic order parameter $\Phi$. 

\begin{figure}
\includegraphics[width=0.99\linewidth]{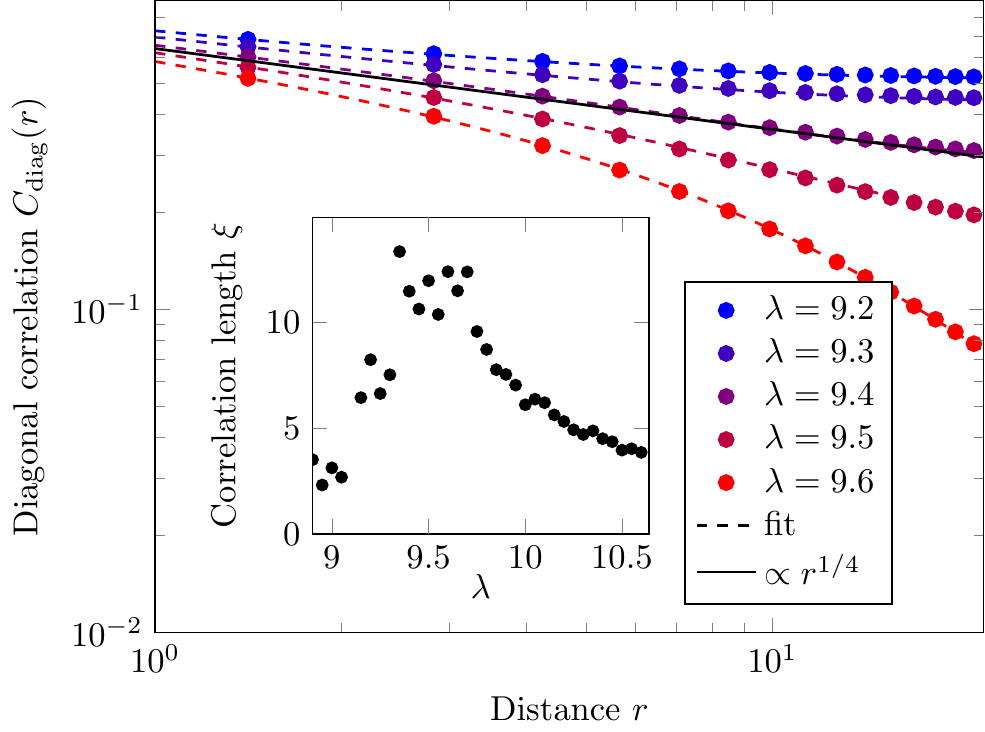}
\caption{\label{fig: correlation}Diagonal correlation function $C_\mathrm{diag}(r)$ for different $\lambda$. Below and above $\lambda_\mathrm{c} \approx 9.4$, $C_\mathrm{diag}(r)$ decays exponentially to zero or $C_0 = \langle m_\pm\rangle^2$, respectively. At $\lambda_\mathrm{c}$, we observe power-law behavior with an exponent $\eta \approx 1/4$ (slope shown as solid black line). Inset: correlation length $\xi$ as a function of $\lambda$.}
\end{figure}

\paragraph*{Correlation function.}
To further characterize the second-order transition apparent from the order parameter, we consider the 2D spatial correlation function
\begin{equation}
\label{eq: correlation function sub}
C(n,m) = \frac{2}{N}\left\langle \sum_{i}^{n_x}\sum_{j}^{n_y} S_{i,j}S_{i+n,j+m} \right\rangle\, ,
\end{equation}
where $n$ and $m$ are integer steps.
In an antiferromagnetic lattice, the full 2D correlation corresponds to a chessboard pattern, see SM.
For our purposes, it is sufficient to look at the diagonal ($n=m$) correlation function $C_\mathrm{diag}(r)$, which is only a function of distance, $r=\sqrt{n^2+m^2}$.
In Fig~\ref{fig: correlation}, $C_\mathrm{diag}(r)$ is plotted in a log-log scale at different values of $\lambda$ for the largest system size of $64\times 64$ vortices.
The correlation function decays exponentially to zero for $\lambda > \lambda_\mathrm{c}$, whereas for $\lambda < \lambda_\mathrm{c}$, it reaches a finite value $C_0 = \langle m_\pm\rangle^2$.
We fit the correlation function via $C_\mathrm{diag}(r) - C_0 = A_C \, r^{2-d-\eta} \, e^{-r/\xi}$~\cite{henkel1999conformal,sethna2005statistical}, where $\xi$ is the correlation length, $A_C$ is an additional fitting parameter and $d=2$ the dimension of the system.
At $\lambda_\mathrm{c}\approx 9.4$, we observe power-law behavior with an exponent $\eta \approx 1/4$, which is shown as a solid line in Fig~\ref{fig: correlation}.
The correlation length $\xi$ is shown as an inset, indicating divergent behavior.
The statistics of $\xi$ does not allow to extract reliably a critical exponent; this would require an extensive finite-size scaling analysis which is outside the scope of the present study.
Still, the overall behavior of $C_\mathrm{diag}(r)$, particularly the exponent $ \eta = 1/4$, is reminiscent of a second-order transition in the universality class of the 2D Ising model with nearest-neighbor interactions~\cite{henkel1999conformal}.
This picture is further supported by data of the susceptibility below $\lambda_\mathrm{c}$ (see SM), showing power-law behavior with exponent $\gamma = 7/4$.
Further, as expected close to a second-order transition, we observe \textit{critical slowing down}, 
i.e., a profound increase of the extent of temporal correlations of $\Phi(t)$ upon
approaching $\lambda_\mathrm{c}$ (see SM).

\paragraph*{Effective temperature.} 

Given the "Ising-like" behavior of the order parameter (and the other quantities studied here) as functions of the strength of nonlinear advection, $\lambda$, it is an intriguing question whether we can relate $\lambda$ to an \textit{effective} temperature $T_\mathrm{eff}$ of our non-equilibrium system. 
As a starting point, we set $T_\mathrm{eff}$ equal to $k_\mathrm{B}T/J$ entering Eq.~(\ref{eq: Onsager's solution}), solve this equation with respect to $T_\mathrm{eff}$ and take the numerical result for $\Phi(\lambda)$  (see Fig.~\ref{fig: order}) to calculate $T_\mathrm{eff}$ as a function of $\lambda$.
Clearly, this can only be done in the range $\Phi \neq 0$, i.e., $\lambda < \lambda_\mathrm{c}$.
The result is shown in the inset of Fig.~\ref{fig: order}a).
Remarkably, in the range of $\lambda$-values where spin fluctuations occur ($\lambda>5$), we find a \textit{linear} dependence between $T_\mathrm{eff}$ and $\lambda$, specifically $T_\mathrm{eff}(\lambda) \approx 0.313(\lambda - 2.05)$, suggesting that we can indeed consider $\lambda$ as an effective temperature (up to some shift).
This correspondence is shown in Fig.~\ref{fig: order}a) by the second x-axis.
Note that the deviation from Onsager's analytical solution above the critical temperature is expected due to the finite system size.
From the relation $T_\mathrm{eff}(\lambda)$, it follows that absolute zero, i.e., $T_\mathrm{eff} = 0$, corresponds to $\lambda^\star = 2.05$.
Remarkably, this value coincides with $\lambda^\star \approx 2$, above which the square lattice ``ground state'' in the unconstrained system becomes unstable to the formation of a dynamic, mesoscale-turbulent state.
A further intriguing consequence of the linearity $T_\mathrm{eff} \propto \lambda$ appears when we relate $\lambda$ to microscopic parameters (see 
SM), particular the self-swimming speed $v_0$ that is experimentally tunable~\cite{sokolov2012physical}.
In fact, we find $T_\mathrm{eff} \propto \lambda\propto v_0$, in contrast to studies of spherical active particles where $T_{\mathrm{eff}}\propto D\propto v_0^2$ \cite{romanczuk2012active,cates2015motility} (with $D$ being the diffusion constant).
We further note that the existence of a linear relation $T_\mathrm{eff}(\lambda)$ is robust against details of the setup, which only lead to changes of the quantitative mapping $\lambda \rightarrow T_\mathrm{eff}$.
The scenarios tested include varying obstacle size $\ell$ and lattice constant $L$ as well as introducing a small amount of disorder into the system by randomly removing a few obstacles, see SM.
For example, changing $\ell = 0.13 \Lambda$ to $\ell = 0.11 \Lambda$ yields $T_\mathrm{eff}(\lambda) \approx 0.46(\lambda - 2.08)$.

\paragraph*{Conclusions}
While antiferromagnetic vortex structures in active fluids are now well established~\cite{wioland2013confinement,lushi14fluid,wioland2016ferromagnetic,reinken2020organizing}, the present study substantially broadens the picture: By considering the strength of nonlinear advection, $\lambda$, as a tunable parameter, we 
found that the vortex lattice transforms via a second-order phase transition with Ising-like characteristics into 
a disordered state, namely, mesoscale turbulence. 
At the critical point, the range of spin-spin correlations (i.e.. correlations between vortex rotation) diverges, indicating
pattern formation on much larger scales. Our analysis moreover reveals the presence of an effective temperature
directly proportional to $\lambda$, quite different from earlier studies of non-equilibrium systems, where effective temperatures have been defined~\cite{loi2008effective,palacci2010sedimentation,cugliandolo2011effective,maggi2014generalized,takatori2015theory}.
From a more general perspective, our study complements recent attempts to relate complex non-equilibrium transitions to (standard) models from statistical physics,
other prominent examples being the onset of turbulence in inertial fluids~\cite{lemoult2016directed} and active nematics~\cite{doostmohammadi2017onset} viewed as directed percolation. 
We note that, apart from a different geometry, here we also focused on a different order parameter.
Clearly, further work is necessary to elucidate such connections and their
practical relevance for the engineering of active and biological fluids.

\appendix

\begin{acknowledgments}
This work was funded by the Deutsche Forschungsgemeinschaft (DFG, German Research Foundation) - Projektnummer 163436311 - SFB 910.
SH and MB acknowledge support by the Deutsche Forschungsgemeinschaft (DFG) through Grants HE 5995/3-1 (SH), BA 1222/7-1 (MB) and SFB 910. 
\end{acknowledgments}

\input{main.bbl}

\end{document}


\title{Ising-like critical dynamics of vortex lattices in an active fluid\\[1\baselineskip] \textit{Supplemental material}}

\author{Henning Reinken}
\email{henning.reinken@itp.tu-berlin.de}
\affiliation{Technische Universit\"at Berlin, Institute of Theoretical Physics, Stra{\ss}e des 17. Juni 135, 10623 Berlin, Germany} 

\author{Sebastian Heidenreich}
\affiliation{Department of Mathematical Modelling and Data Analysis,
Physikalisch-Technische Bundesanstalt Braunschweig und Berlin, Abbestr. 2-12,
10587 Berlin, Germany}

\author{Markus B\"ar}
\affiliation{Department of Mathematical Modelling and Data Analysis,
Physikalisch-Technische Bundesanstalt Braunschweig und Berlin, Abbestr. 2-12,
10587 Berlin, Germany}

\author{Sabine H. L. Klapp}
\affiliation{Technische Universit\"at Berlin, Institute of Theoretical Physics, Stra{\ss}e des 17. Juni 135, 10623 Berlin, Germany} 

\date{\today}

\maketitle

This supplemental material contains information on the methods used in this work and additional plots and discussion on topics touched upon in the main text.
In the first section, we introduce the numerical methods, discuss initial conditions, the averaging procedure and calculation of error bars.
Then, we briefly summarize the dynamics in the unconstrained system, discuss the full 2D spatial correlation function, the definition of the effective susceptibility and its critical behavior.
After a discussion on the symmetry properties of the system, we analyze the dynamics and show that the system exhibits critical slowing down.
Further, we explore how the macroscopic coefficients in our model depend on microscopic parameters like the swimming speed.
In addition, we discuss the impact of slight changes of the properties of the obstacle arrays, including changing the obstacle size and the lattice constant and removing a small number of obstacles from the lattice.

\section{Numerical methods}

\subsection*{Continuum equation in the presence of obstacles}

Evaluating the functional derivative, Eq.~(1) for the effective microswimmer velocity $\mathbf{v}$ in the main text can be written as
\begin{equation}
\label{eq: dynamic equation non potential}
\partial_t \mathbf{v} + \lambda \mathbf{v}\cdot\nabla\mathbf{v} = - \nabla q + a \mathbf{v} - b |\mathbf{v}|^2\mathbf{v} - (1 + \nabla^2)^2 \mathbf{v}.
\end{equation}
We further reformulate Eq.~(\ref{eq: dynamic equation non potential}) in terms of the vorticity $\omega = (\nabla\times {\bf v})_z$ and incorporate the effects of obstacles via
\begin{equation}
\label{eq: vorticity formulation}
\partial_t \omega + \lambda \mathbf{v}\cdot\nabla\omega = a \omega  - b \big[\nabla \times (|\mathbf{v}|^2\mathbf{v})\big]_z   - (1 + \nabla^2)^2 \omega
- \gamma_\mathbf{v} \big\{\nabla\times[K(\mathbf{r})\mathbf{v}]\big\}_z - \gamma_\omega K(\mathbf{r}) \omega.
\end{equation}
Recent experiments~\cite{nishiguchi2018engineering,reinken2020organizing} have shown that small obstacles locally reduce both, the velocity $\mathbf{v}$ and the vorticity $\omega = (\nabla \times \mathbf{v})_z$.
These boundary conditions, i.e., $\mathbf{v} = 0$ and $\omega = 0$, are incorporated by a local damping potential, specified by the two additional terms in Eq.~(\ref{eq: vorticity formulation}) with the damping coefficients $\gamma_\mathbf{v}$ and $\gamma_\omega$, respectively.
The kernel (or ``mask'') $K(\mathbf{r})$ contains all the information about the shape, geometry and configuration of obstacles by specifying the damping at position $\mathbf{r}$.
Here, we choose the following generalized Gaussian function
\begin{equation}
\label{eq: kernel generalized gaussian}
K(\mathbf{r}) = \sum_i e^{\ln(0.0001) \big( \frac{\mathbf{r} - \mathbf{r}_i}{\ell /2} \big)^4 },
\end{equation}
where $\mathbf{r}_i$ is the location of obstacle $i$.
The obstacle diameter $\ell$ is defined by the distance to the center where the damping is reduced to $\SI{0.01}{\percent}$ of $\gamma_\mathbf{v}$ and $\gamma_\omega$, respectively.
This approach has proven to be successful in describing the effect obstacles have on the surrounding flow field in terms of topological charges centered in the obstacles~\cite{reinken2020organizing}.
In particular, we are able to incorporate the effect of increasing the obstacle size, which leads to more negative charges, i.e., more surrounding vortices.
This finding has also been corroborated by experiments~\cite{reinken2020organizing}.
Further, this approach was shown to capture the emergence of antiferromagnetic order in lattices of different symmetry~\cite{reinken2020organizing} and reproduces experiments with a bacterial suspension of \textit{Bacillus Subtilis} in a square lattice configuration of obstacles~\cite{nishiguchi2018engineering}.
In this work, we choose the same obstacle size that has been used in the experiments, i.e, $\ell = 0.13 \Lambda$.
This size corresponds to a preference for four surrounding vortices around each obstacle~\cite{reinken2020organizing}, consistent with the geometry of a square lattice.
Further, we set $a=0.5$, $b=1.6$, $\gamma_\mathbf{v} = 40$ and $\gamma_\omega = 4$, which reproduces experimental results on the effect of a small 3D-printed pillar on \textit{Bacillus Subtilis} suspensions~\cite{reinken2020organizing}.

\subsection*{Numerical solution}

Equation~(\ref{eq: vorticity formulation}) is solved with a pseudospectral method.
Using operator splitting, the linear part is solved exactly and the nonlinear part is integrated in time applying Euler's method.
In contrast to numerical computations of classic spin models such as the Ising model (e.g., using Monte Carlo methods), we have to resolve every spin spatially, which makes our calculations quite costly.
In this work, we use a resolution of $32\times32$ grid points for every gap between obstacles.
We tested different resolutions and found this size to be large enough to avoid numerical artifacts.
As a result, the total spatial resolution is up to $2048\times 2048$ grid points for the largest system consisting of $64 \times 64$ obstacles.

\subsection*{Initial conditions}

To check whether the antiferromagnetic order emerges indeed spontaneously (at appropriate value of $\lambda$), we first set the initial conditions to the quiescent state, i.e.,  $\mathbf{v} = 0$, and add random local perturbations.
After this test, we perform all subsequent numerical calculations starting from already ordered initial conditions, i.e., from an antiferromagnetic lattice with additional local perturbations.
This approach allows to avoid the otherwise very long relaxation times.
Before starting with the analysis of the vortex patterns, we let the system evolve for $1000$ time units, which is at least an order of magnitude larger than it takes the unconstrained system to develop mesoscale turbulence from a quiescent initial state.

\subsection*{Averaging procedure}

To calculate averaged quantities, such as the order parameter $\Phi$ and the correlation function $C_\mathrm{diag}(r)$, we perform computational runs with different initial perturbations, take the temporal average for every run and additionally average over all the computational runs.
The closer we are to the critical point, the more sampling we need in order to obtain accurate statistics.
For some values of $\lambda$, the effective averaging time adds up to more than $100,000$ time units.

\subsection*{Error bars}

We include error bars in the plots for quantities that are directly calculated from the averaging procedure, e.g., the order parameter $\Phi$, the Binder cumulant $U$ and the susceptibility $\chi$.
These error bars represent the standard error, i.e., the standard deviation of the respective quantity divided by the square root of the number of runs.

\clearpage

\section{Dynamics in the unconstrained system}

\begin{figure}[h]
\includegraphics[width=0.99\linewidth]{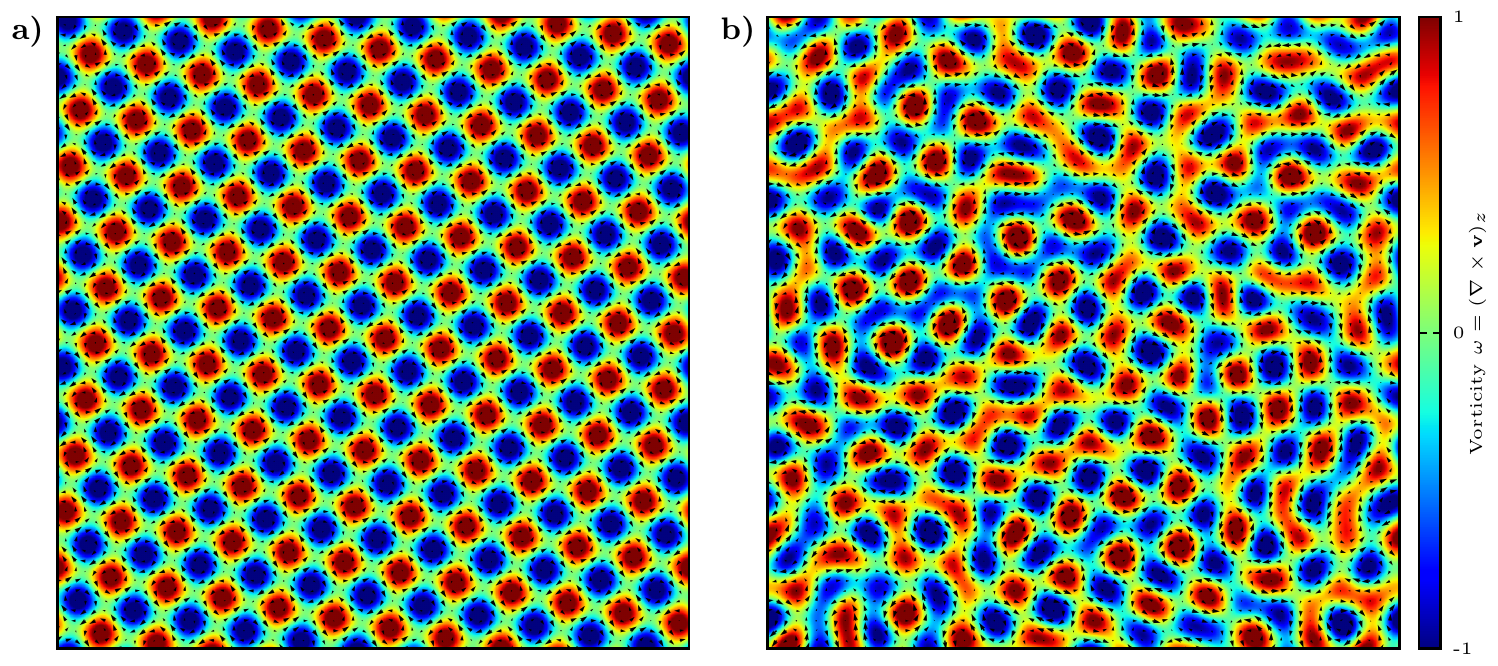}
\caption{\label{fig: snapshots unconstrained}Snapshots of the vorticity field $\omega$ for \textbf{a)} $\lambda = 1.9 < \lambda^\star$ and \textbf{b)} $\lambda = 2.1 > \lambda^\star$. The arrows denote the velocity field $\mathbf{v}$. The initial conditions for theses calculations were set to the quiescent state, i.e., $\mathbf{v} = 0$, with additional small local perturbations. The other parameters are $a = 0.5$ and $b=1.6$.}
\end{figure}

\begin{figure}[h]
\includegraphics[width=0.55\linewidth]{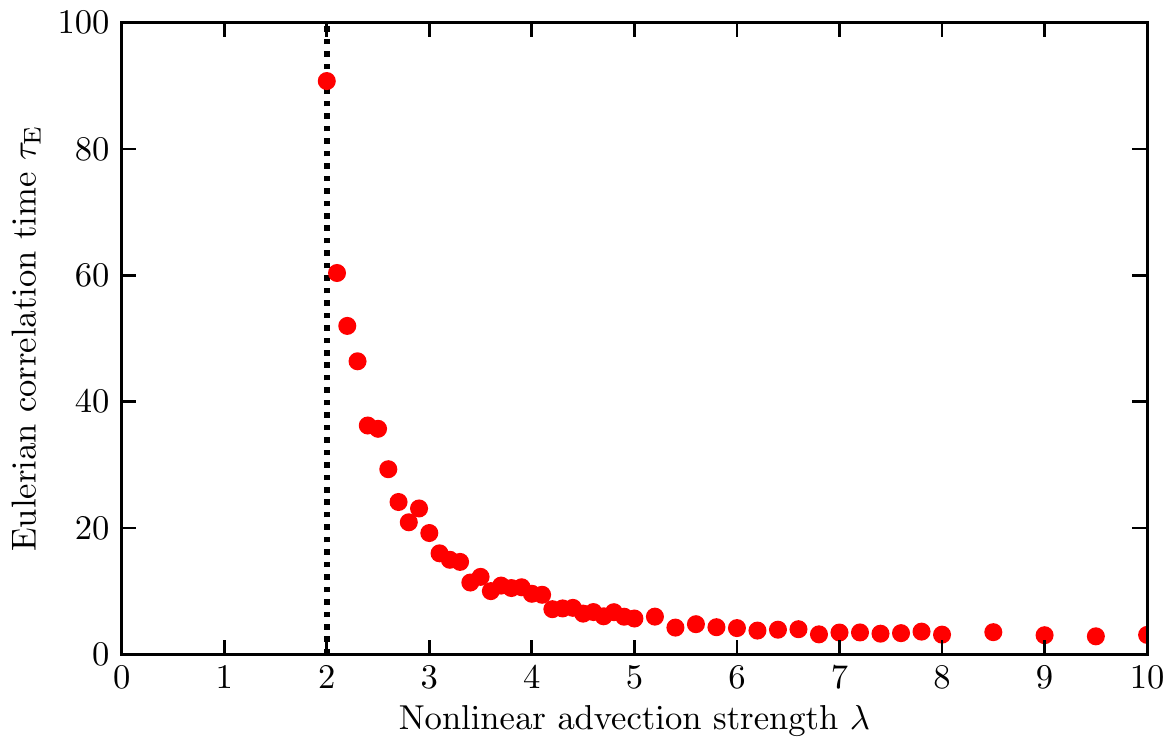}
\caption{\label{fig: correlation time unconstrained}Eulerian correlation time $\tau$ as a function of nonlinear advection strengh $\lambda$ in the unconstrained system. The dotted line at $\lambda \approx 2$ marks the value below which the system settles into a stationary vortex lattice when starting from an initially quiescent state $\mathbf{v} = 0$. At this point, $\tau$ diverges. At larger values of $\lambda$, the system develops a dynamic state, i.e., the correlation times becomes finite. The other parameters are $a = 0.5$ and $b=1.6$.}
\end{figure}

The dynamics  of the flow field $\mathbf{v}$ in the unconstrained system depends strongly on the strength of nonlinear advection $\lambda$ (see Eq.~(1) in the main text).
Starting from an initially quiescent state, i.e., $\mathbf{v} = 0$, a square lattice of vortices emerges, provided $\lambda$ is below some critical value $\lambda^\star$, which depends on the other parameters $a$ and $b$.
The length scale is determined by $\Lambda = 2 \pi / k_\mathrm{c}$, where $k_\mathrm{c}$ is the wavenumber of the fastest-growing mode (see~\cite{dunkel2013minimal,heidenreich2016hydrodynamic,reinken2018derivation,reinken2019anisotropic} for details).
This square vortex lattice state has been shown to be a minimum of the potential functional $\mathcal{F}$ introduced in Eq.~(1) in the main text, see also~\cite{james2018turbulence,reinken2019anisotropic}.
In Fig.~\ref{fig: snapshots unconstrained}a), a snapshot of the vorticity field $\omega$ in the square lattice state is shown.
Provided $\lambda < \lambda^\star$, which means that the dynamics is dominated by the relaxational term determined by $\mathcal{F}$, the system always develops this stationary square lattice state after some relaxtion time.
Increasing $\lambda$ above $\lambda^\star \approx 2$, the nonlinear advection term $\lambda \mathbf{v}\cdot\nabla\mathbf{v}$ [see Eq.~(1) in the main text] induces instabilities and the system develops a dynamic state that has been denoted as mesoscale turbulence~\cite{wensink2012meso}.
This state still exhibits a dominating length scale (or vortex size) close to $\Lambda$.
Fig.~\ref{fig: snapshots unconstrained}b) shows a snapshot of the vorticity field at $\lambda = 2.1$, i.e., right above $\lambda^\star$.
In contrast to the ordered vortex lattice in the system subject to obstacles, however, the vortices are not spatially pinned in the unconstrained case.
Therefore, we can not directly calculate the same quantities (sublattice magnetization, susceptibility, ...) as for the transition in the obstacle arrangement discussed in the main text.
Instead, we characterize the transition from a stationary vortex lattice to a turbulent state in the unconstrained system by the temporal correlations.
To this end, we calculate the Eulerian two-time correlation function $C_\mathrm{E}(\Delta t)$ via
\begin{equation}
\label{eq: Eulerian temporal correlation function}
C_\mathrm{E}(\Delta t) = \big\langle \mathbf{v}(\mathbf{x},t)\cdot\mathbf{v}(\mathbf{x},t + \Delta t)\big\rangle \, ,
\end{equation}
where $\langle \dots \rangle$ denotes an average over both time $t$ and space $\mathbf{x}$.
We define the Eulerian correlation time $\tau_\mathrm{E}$ via the integral
\begin{equation}
\label{eq: Eulerian correlation time}
\tau_\mathrm{E} =\int\limits_0^\infty C_\mathrm{E}(\Delta t) \, d \Delta t \, .
\end{equation}
Fig.~(\ref{fig: correlation time unconstrained}) shows $\tau_\mathrm{E}$ plotted as a function of nonlinear advection strength $\lambda$.
Above the critical value $\lambda^\star \approx 2$, the unconstrained system develops a dynamic, turbulent state, temporal correlations decay to zero and the correlation time $\tau_\mathrm{E}$ is finite.
Increasing $\lambda$ leads to a decrease in the correlation time $\tau_\mathrm{E}$, i.e., to faster changes of the structures.
In contrast, below $\lambda^\star$, the system settles into a stationary vortex lattice, i.e., the structures are frozen in time, which leads to a diverging correlation time $\tau_\mathrm{E}$ at this point, as is visible in Fig.~(\ref{fig: correlation time unconstrained}). 
There are no fluctuations in this ``ground state'' below $\lambda^\star \approx 2$ and therefore also no concept of temperature.


\clearpage

\section{Two-dimensional spatial correlation function}

\begin{figure}[h]
\includegraphics[width=0.99\linewidth]{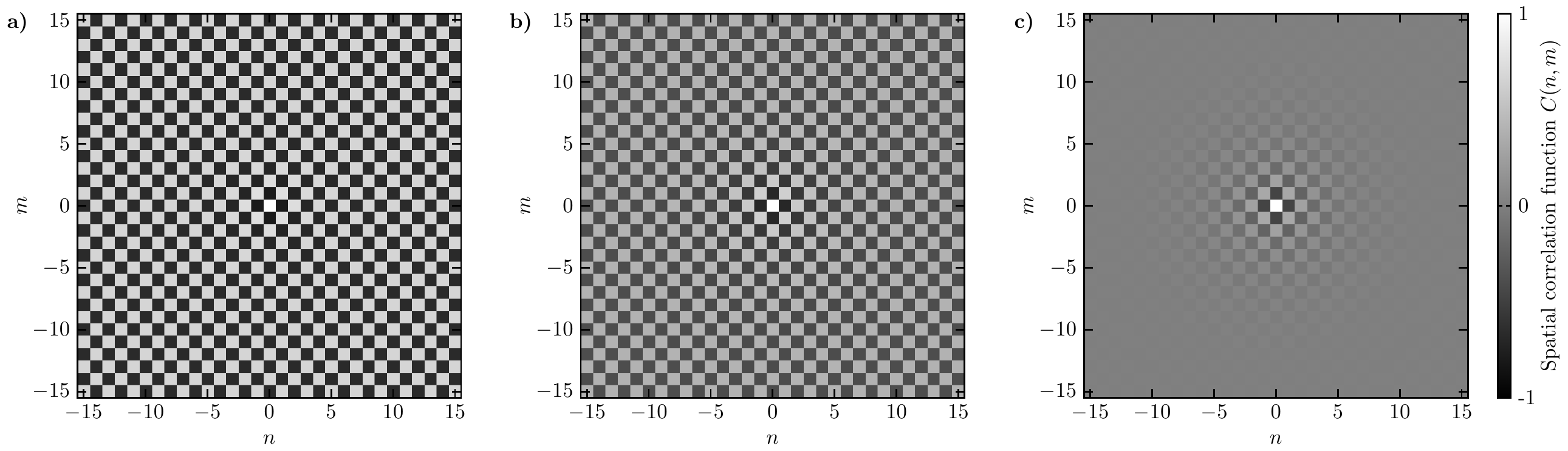}
\caption{\label{fig: full spatial corr}Two-dimensional spatial correlaton function $C(n,m)$ \textbf{a)} in the ordered state at $\lambda = 9$, \textbf{b)} near the critical point at $\lambda = 9.4$ and \textbf{c)} in the disordered state at $\lambda = 10.9$. The system size is $N=32\times 32$.}
\end{figure}

In the main text, we define the full spatial correlation function $C(n,m)$, see Eq.~(4), which is both dependent on the x- and y-direction.
For the sake of simplicity, we calculate the correlation length from the diagonal correlation $C_\mathrm{diag}(r)$, which is only dependent on the distance $r = \sqrt{n^2+m^2}$, see main text.
To further illustrate the behavior at the transition, we show the full 2D correlation function $C(n,m)$ in Fig.~\ref{fig: full spatial corr} for different values of nonlinear advection strength $\lambda$.
In part a), where we are below the critical point in the ordered state ($\lambda = 9$), we observe a chessboard pattern characteristic for the long-range antiferromagnetic order.
The spatial correlations decrease when increasing $\lambda$.
In part b), close to the critical point $\lambda_\mathrm{c} = 9.4$, the system is still correlated for the full extent of its finite size but the average magnetization is lower than in part a).
Finally, in the disordered state, the spins decorrelate after a short distance, which is visible as ``graying-out'' in Fig.~\ref{fig: full spatial corr}c), where the correlation function at $\lambda = 10.9$ is shown.

\clearpage

\section{Definition and critical scaling of the effective susceptibility}

In this work, we define the effective sublattice susceptibility $\tilde{\chi}_\pm$ in terms of the fluctuations of the magnetization $M_\pm$, i.e, its variance,
\begin{equation}
\label{eq: susceptibility}
\tilde{\chi}_\pm =  \langle M_\pm^2\rangle - \langle M_\pm\rangle^2 \, ,
\end{equation}
where $M_\pm$ is defined as
\begin{equation}
\label{eq: magnetization}
M_\pm = \sum_{i=1}^{n_x} \sum_{j=1}^{n_y} \left(1\pm(-1)^{(i+j)}\right) S_{i,j} \, .
\end{equation}
We here omit the prefactor $(k_\mathrm{B}T)^{-1}$ in Eq.~(\ref{eq: susceptibility}) often used in the context of magnetic systems~\cite{odor2004universality}.
Note that this omission does not crucially impact the scaling behavior near the critical point as $\tilde{\chi}_\pm$  diverges with $|T - T_\mathrm{c}|$ rather than $T$.
The effective susceptibility $\tilde{\chi}_\pm$ is an extensive quantity, i.e., it scales linear in $N$ when increasing the system size.
In order to obtain an intensive quantity, we divide by $N/2$, which yields the susceptibility per lattice site, $\chi_\pm$,
\begin{equation}
\label{eq: susceptibility per lattice site}
\chi_\pm = \frac{\tilde{\chi}_\pm}{N/2} = \frac{\langle M_\pm^2\rangle - \langle M_\pm\rangle^2}{N/2} =  \frac{N}{2} \big(\langle m_\pm^2\rangle - \langle m_\pm\rangle^2 \big)\, ,
\end{equation}
where $m_\pm = M_\pm/(N/2)$ is the sublattice magnetization per lattice site.

\begin{figure}[b]
\includegraphics[width=0.5\linewidth]{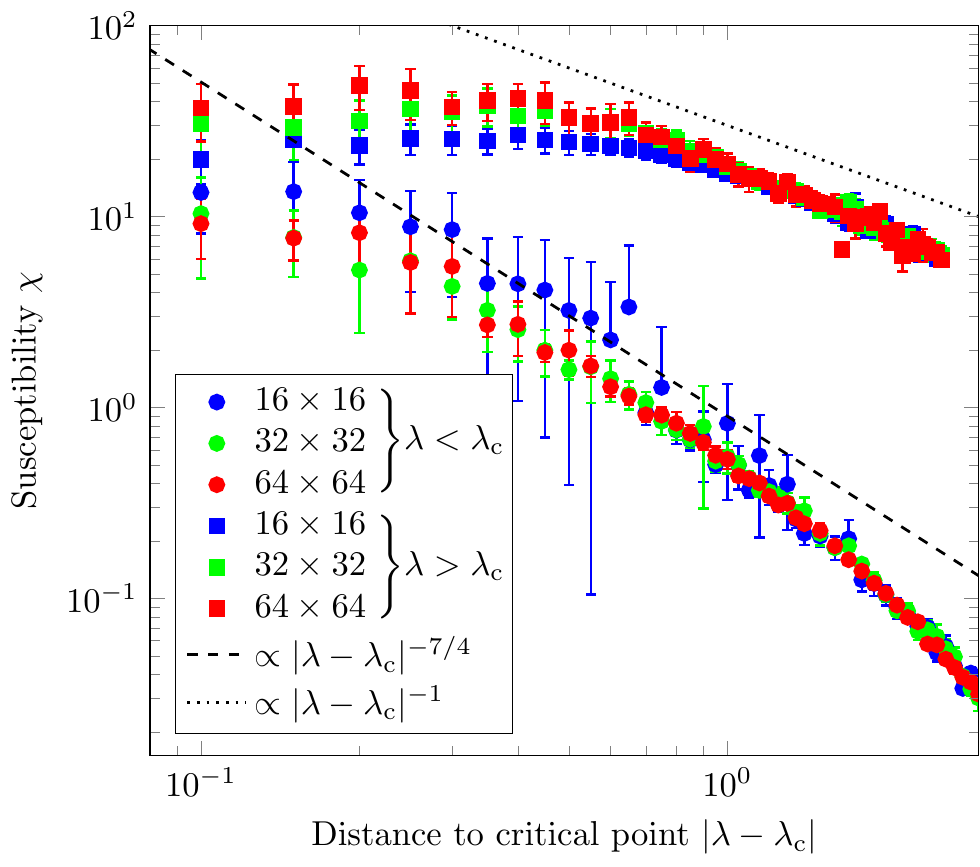}
\caption{\label{fig: susceptibility}Susceptibility $\chi$ as a function of the distance to the critical point $|\lambda-\lambda_\mathrm{c}|$ in a log-log plot for different system sizes $N=16$, $N=32$ and $N=64$. Below the critical point, the divergence of the susceptibility is consistent with a power law with an exponent $\gamma\approx 7/4$ (slope shown as dashed line), which corresponds to the 2D Ising universality class. Above the critical, however, we observe $\gamma = 1$ (slope shown as dotted line), which is consistent with the Curie-Weiss law. The error bars represent the standard error.}
\end{figure}

To further corroborate the result that the antiferromagnetic transition in the bacterial vortex lattice falls into the 2D Ising universality class, analyze the critical behavior of the susceptibility $\xi$, which is plotted in Fig.~\ref{fig: susceptibility} as a function of the distance to the critical point $|\lambda - \lambda_\mathrm{c}|$.
For systems belonging to the 2D Ising universality class, we expect the susceptibility to diverge when approaching the critical point with an exponent $\gamma = 7/4$, i.e., $\chi_\pm \propto |T - T_\mathrm{c}|^{-\gamma}$.
Having established the linear dependence $T_\mathrm{eff}(\lambda)$, we expect $\chi_\pm \propto |\lambda - \lambda_\mathrm{c}|^{-\gamma}$ as well.
Fig.~\ref{fig: susceptibility} shows the susceptibility $\chi$ (averaged over the sublattices) as a function of the distance to the critical point $|\lambda- \lambda_\mathrm{c}|$ on a log-log scale.
When approaching the critical point from below (lower data set), the data points are consistent with power-law behavior characterized by an exponent $\gamma\approx 7/4$ expected for the 2D Ising universality class, even for the small to medium-sized systems considered in this work.
In contrast, $\gamma \approx 1$ far above $\lambda_\mathrm{c}$ (upper data set), indicating Curie-Weiss (mean-field) behavior.

\clearpage

\section{Discussion on symmetry properties and continuous spins}

\begin{figure}[h]
\includegraphics[width=0.9\linewidth]{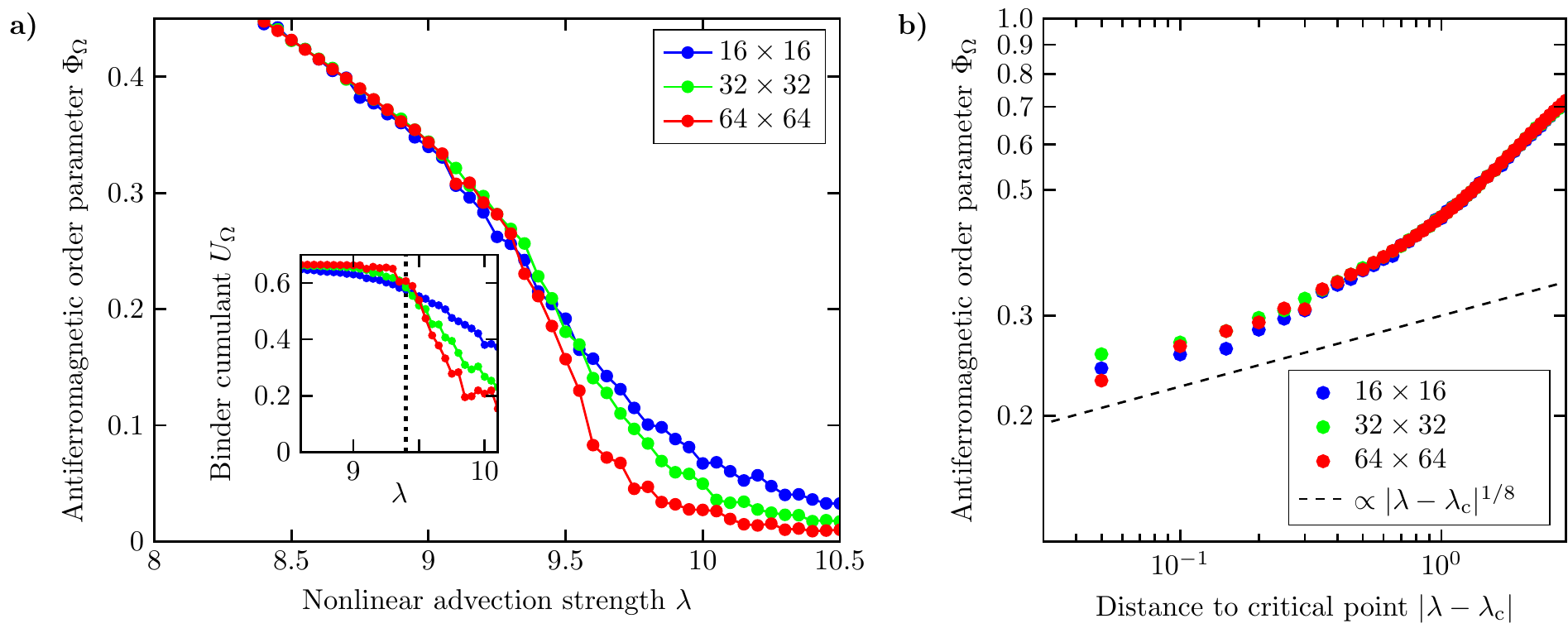}
\caption{\label{fig: order vort}\textbf{a)} Antiferromagnetic order parameter $\Phi_\Omega$ calculated from gap-averaged vorticities $\Omega_{i,j}$ for different system sizes $N=16\times 16$, $N=32\times 32$ and $N=64\times 64$. In \textbf{a)}, $\Phi_\Omega$ is plotted as a function of the nonlinear advection strength $\lambda$. Inset: Binder cumulant $U_\Omega$ as a function of $\lambda$. In \textbf{b)}, $\Phi_\Omega$ is shown as a function of the distance to the critical point $|\lambda - \lambda_\mathrm{c}|$ with $\lambda_\mathrm{c} = 9.4$ on a log-log scale. Close to the critical point we observe behavior consistent with 2D Ising universality class, i.e, the order parameter grows as $\Phi_\Omega \propto |\lambda - \lambda_\mathrm{c}|^\beta$, where $\beta = 1/8$.}
\end{figure}

Another possibility to map the emerging dynamics of the full vorticity field to a spin system would be to directly
use the continuous variables $\Omega_{ij}$ (defined in Eq.~(2) in the main text as the gap-averaged vorticity) instead of the
discrete spins $S_{ij}$. Thus, the variable assigned to each lattice site is still scalar, but takes continuous values. Still, the symmetry of the resulting structures against sign reversal of all variables remains preserved. One might therefore speculate that the qualitative behavior at the transition from disorder to an ordered antiferromagnetic vortex lattice does not
depend on whether we choose continuous or discrete variables (as in discrete or continuous Ising models~\cite{bruce1985universality,bialas2000percolation})).
To test this conjecture,
we analyze the dynamics of the system consisting of the gap-averaged vorticities $\Omega_{i,j}$ as well.
Analogously to the discrete spin system, we define sublattice magnetizations $m^\Omega_\pm$ via
\begin{equation}
\label{eq: magnetization Omega}
m^\Omega_\pm = \frac{1}{N}\sum_{i=1}^{n_x} \sum_{j=1}^{n_y} \left(1\pm(-1)^{(i+j)}\right) \Omega_{i,j} \, .
\end{equation}
The antiferromagnetic order parameter $\Phi_\Omega$, obtained via $\Phi_\Omega = | \langle m^\Omega_{+} \rangle - \langle m^\Omega_{-}\rangle | /2 $, is shown in Fig.~\ref{fig: order vort}a) as a function of the nonlinear advection strength $\lambda$.
As for the discrete spins, we observe a transition from an ordered state to a disordered state when increasing $\lambda$.
To determine the critical point, we calculate the Binder cumulant $U_\Omega$ analogous to the definition in the main text.
The intersection point of the curves $U_\Omega(\lambda)$ for different system sizes, yields the critical point, $\lambda_\mathrm{c} \approx 9.4$, see inset of Fig.~\ref{fig: order vort}a), which is consistent with the value determined from the discrete-spin analysis.
In Fig.~\ref{fig: order vort}b) we show the antiferromagnetic order parameter $\Phi_\Omega$ as a function of the distance to the critical point $|\lambda - \lambda_\mathrm{c}|$.
For a phase transition belonging to the 2D Ising universality class, we expect the order parameter to grow as $\Phi_\Omega \propto |\lambda - \lambda_\mathrm{c}|^\beta$ where $\beta = 1/8$ close to the critical point.
This is shown as a guide to the eye in Fig.~\ref{fig: order vort}b).
Close to the critical point $\Phi_\Omega$ exhibits behavior consistent with the 2D Ising universality class.
Comparing the slope in Fig.~\ref{fig: order vort}b) with Fig.~2c) in the main text, where the discrete-spin order parameter is shown, we note that the choice between continuous variables and discrete spins does matter once we move further away from the critical point.
In this work, we choose discrete spins for the analysis.
In fact, this seems to be the more natural choice, because the order parameter $\Phi$ (calculated from discrete spins) only incorporates the antiferromagnetic ordering and, in contrast to $\Phi_\Omega$ (calculated from the gap-averaged vorticities), is independent of the strength of the vortices.

\clearpage

\section{Critical slowing down}

\begin{figure}[h]
\includegraphics[width=0.9\linewidth]{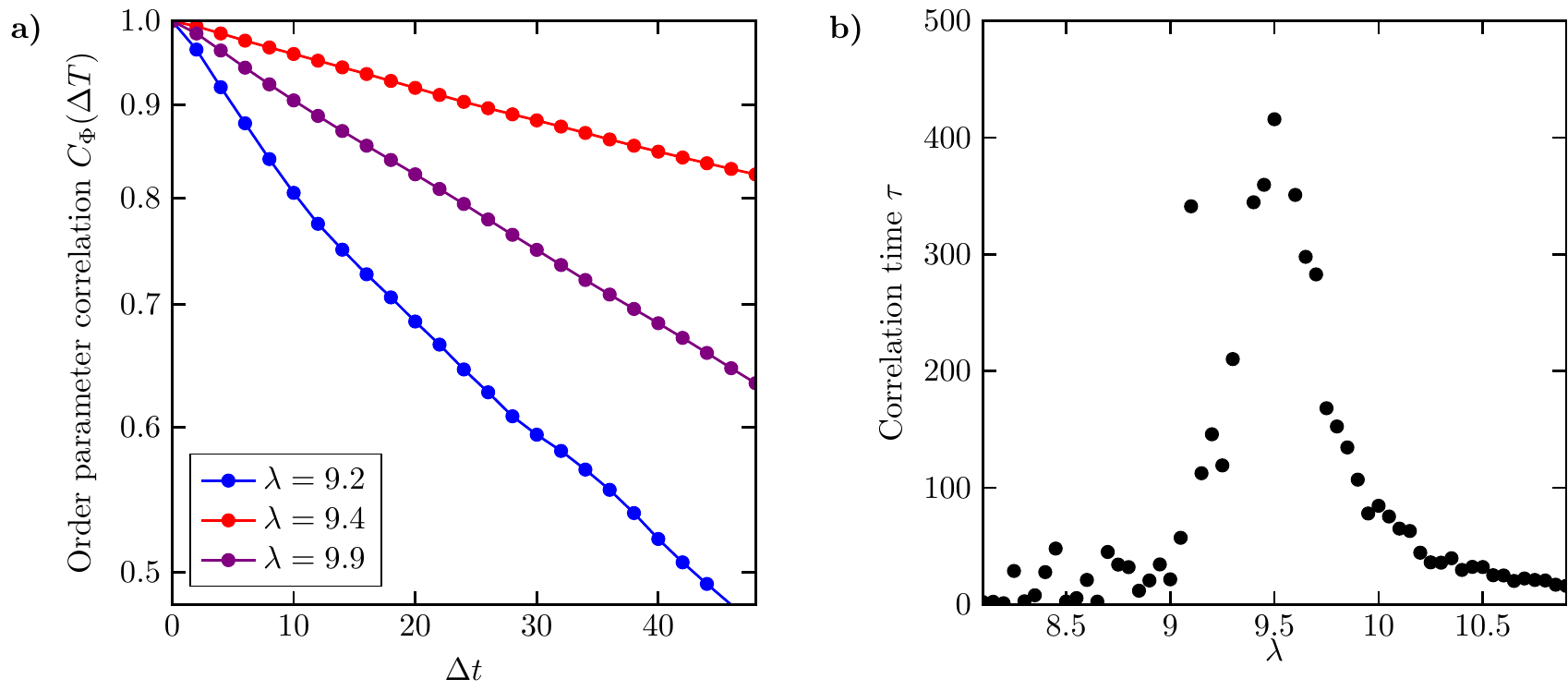}
\caption{\label{fig: temporal correlation}\textbf{a)} Temporal correlations of the instantaneous antiferromagnetic order parameter $\Phi$ for different values of $\lambda$ (using linear scaling for the x-axis and log scaling for the y-axis) showing exponential decay. We observe the slowest decay near the critical point $\lambda_\mathrm{c} \approx 9.4$. \textbf{b)} Correlation time $\tau$ as a function of nonlinear advection strength $\lambda$. We observe critical slowing down, i.e., divergent behavior at the critical point $\lambda_\mathrm{c} \approx 9.4$.}
\end{figure}

Critical slowing down is a phenomenon observed in second-order phase transitions, where the dynamics becomes very slow when approaching the critical point~\cite{hohenberg1977theory}.
In order to investigate the dynamics in our bacterial vortex lattice, we determine the normalized temporal correlation function of the instantaneous antiferromagnetic order parameter, $\Phi_\mathrm{inst.}(t) = |m_{+}(t) - m_{-}(t)|$, calculated via
\begin{equation}
\label{eq: temporal correlation function}
C_\Phi (\Delta t) = \frac{\left\langle \Phi_\mathrm{inst.}(t)\Phi_\mathrm{inst.}(t+\Delta t)\right\rangle - \langle \Phi_\mathrm{inst.}(t) \rangle^2 }{\langle [\Phi_\mathrm{inst.}(t)]^2 \rangle - \langle \Phi_\mathrm{inst.}(t) \rangle^2} \, .
\end{equation}
Fig.~\ref{fig: temporal correlation}a) shows $C_\Phi (\Delta t)$ for different values of $\lambda$ in a semi-log plot.
We infer from the linear slope of the curves at large $\Delta t$ that $C_\Phi (\Delta t)$ decays exponentially, i.e.,
\begin{equation}
\label{eq: correlation time}
C_\Phi (\Delta t) \approx \exp(-\Delta t / \tau) \, ,
\end{equation}
where $\tau$ denotes the correlation time.
Near the critical point, we observe the slowest decay.
Fig.~\ref{fig: temporal correlation}b) shows the correlation time obtained from the correlation function $C_\Phi (\Delta t)$ via fitting to Eq.~(\ref{eq: correlation time}).
When approaching the critical point $\lambda_\mathrm{c}$, we observe divergent behavior of $\tau$.
Thus, the bacterial vortex lattice indeed exhibits critical slowing down close to the transition to antiferromagnetic order.

\clearpage

\section{Dependence on microscopic parameters}

In experiments, it is of course not possible to directly vary the macroscopic coefficients of our continuum model.
Thus, in the following, we want to motivate how changing microscopic parameters might impact the dynamics in such a way that our predictions can be tested experimentally.
In particular, we want to show how the macroscopic coefficients depend on the self-swimming speed $v_0$, because it is a quantity that can be tuned, e.g., by changing the oxygen concentration~\cite{sokolov2012physical}.
 
Eq.~(1) in the main text is a rescaled version of a common continuum model for the dynamics in polar microswimmer suspensions~\cite{dunkel2013fluid,dunkel2013minimal,reinken2018derivation,reinken2019anisotropic}, which reads
\begin{equation}
\label{eq: dynamic equation original}
\partial_t \mathbf{v} + \lambda_0 \mathbf{v}\cdot\nabla\mathbf{v} = - \nabla q - \alpha \mathbf{v} - \beta |\mathbf{v}|^2\mathbf{v} + \Gamma_2 \nabla^2 \mathbf{v} + \Gamma_4 \nabla^4 \mathbf{v} \, , \qquad \nabla \cdot \mathbf{v} = 0 \, .
\end{equation}
In~\cite{reinken2018derivation}, Eq.~(\ref{eq: dynamic equation original}) was derived from a microscopic Langevin model including hydrodynamic interactions arising due to active stresses, which yields the coefficients as functions of dimensionless parameters $P_\mathrm{r}$, $c_\mathrm{I}$ and $c_\mathrm{F}$ via
\begin{equation}
\begin{split}
\label{eq: coefficients original}
\lambda_0 &= \frac{3}{5}c_\mathrm{I}\bigg( 1 + \frac{2}{3}a_0 P_\mathrm{r} 
c_\mathrm{F}\bigg)\, , \qquad \alpha = \left(1 - c_\mathrm{I}\right)/P_\mathrm{r}, 
\qquad \beta = \frac{3}{5}c_\text{I}^2/P_\text{r}\, , \\
\Gamma_2 &= \frac{1}{10}\bigg(\frac{\epsilon}{\ell_0}\bigg)^2 
c_\mathrm{I}/P_\text{r} -\frac{a_0}{15}P_\mathrm{r}c_\mathrm{F}\, ,
\qquad \Gamma_4 = - \frac{a_0}{420}P_\mathrm{r}c_\mathrm{F}\, ,
\end{split}
\end{equation}
where $a_0$ is a shape parameter depending on the length $\ell_0$ and diameter $d$ of the swimmers via
\begin{equation}
\label{eq: shape parameter}
a_0 = \frac{(\ell/d)^2-1}{(\ell/d)^2+1}
\end{equation}
The persistence number $P_\mathrm{r}$ defined via
\begin{equation}
\label{eq: persistence number}
P_\mathrm{r} = v_0\tau/\ell_0
\end{equation}
characterizes the persistence of the active motion given by the self-swimming speed $v_0$ relative to rotational diffusion quantified by the relaxation time $\tau$.
The strength of polar interactions compared to rotational diffusion is given by the dimensionless interaction parameter $c_\mathrm{I}$ defined via
\begin{equation}
\label{eq: interaction parameter}
c_\mathrm{I} = \frac{8}{9}\pi \epsilon^3\tau \rho\gamma_0 v_0   \, ,
\end{equation}
where $\rho$ is the number density of swimmers and $\epsilon$ and $\gamma_0$ are the interaction range and strength, respectively.
The strength of the response of the solvent flow to the activity is characterized by the dimensionless coupling parameter $c_\mathrm{F}$ defined via
\begin{equation}
\label{eq: coupling parameter}
c_\mathrm{F} = \frac{f_0\rho \ell_0^2}{10\mu_\mathrm{eff}v_0}\, ,
\end{equation}
where $\mu_\mathrm{eff}$ is the effective viscosity which incorporates active and passive effects of the swimmers, see~\cite{reinken2018derivation} for details.
The active force can be estimated via $f_0 = 2\pi \mu_0 \ell_0 v_0$~\cite{wolgemuth2008collective,heidenreich2016hydrodynamic}, where $\mu_0$ is the solvent viscosity.
Thus, the coupling parameter can be written as 
\begin{equation}
\label{eq: coupling parameter 2}
c_\mathrm{F} = \frac{\pi}{5}\frac{\mu_0}{\mu_\mathrm{eff}}\rho \ell_0^3\, .
\end{equation}
The dynamics is dominated by the polar order parameter when $c_\mathrm{F}$ is small.
In this regime, the dynamics of the solvent flow described by the Stokes equation does not have to be considered separately and the model reduces to Eq.~(\ref{eq: dynamic equation original}), see~\cite{reinken2018derivation} for details.

\begin{figure}[h]
\includegraphics[width=0.99\linewidth]{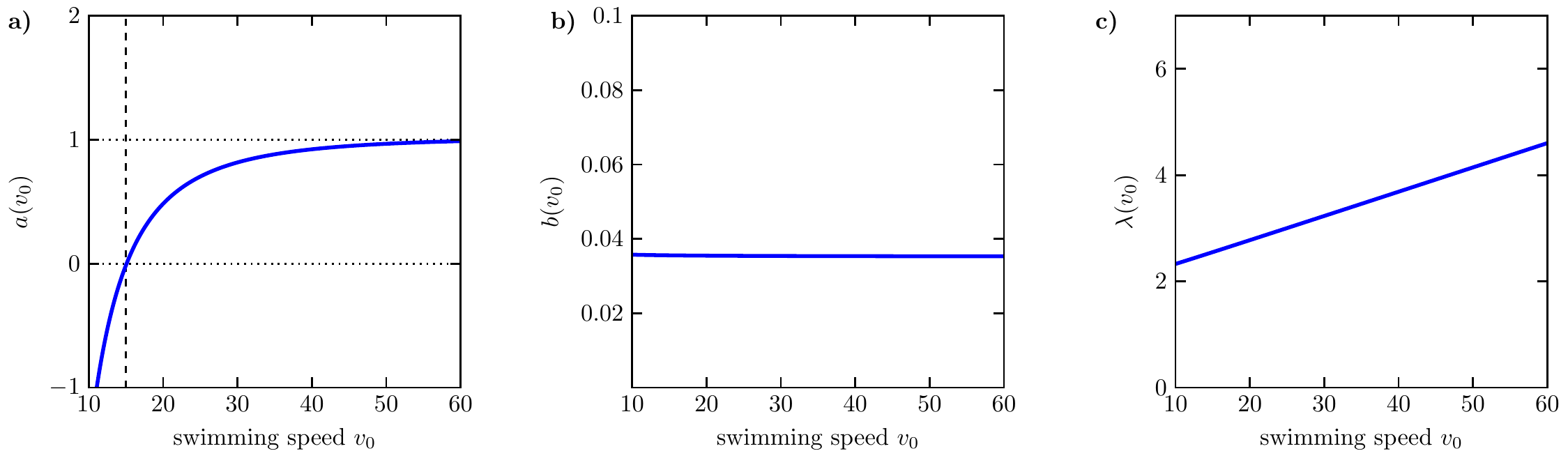}
\caption{\label{fig: swimming speed three coeff} Macroscopic coefficients $a$, $b$ and $\lambda$ as a function of the swimming speed $v_0$ of a single microswimmer. The onset of pattern formation is marked by the dashed line. \textbf{a)} The coefficient $a$ becomes positive at the onset of pattern formation and then increases with $v_0$ approaching a near constant value. \textbf{b)} The coefficient is constant for the whole range of $v_0$. \textbf{c)} The nonlinear advection strength $\lambda$ increases linearly with $v_0$.}
\end{figure}

Rescaling space and time by the time and length scale of the pattern formation, i.e., the inverse of the critical wavenumber $k_\mathrm{c} = \sqrt{\Gamma_2/(2\Gamma_4)}$ and the inverse of the maximum corresponding growth rate $\sigma_\mathrm{c} = -\Gamma_2^2/(4\Gamma_4)$, see e.g.~\cite{reinken2019anisotropic}, yields Eq.~(1) in the main text,
\begin{equation}
\label{eq: dynamic equation}
\partial_t \mathbf{v} + \lambda \mathbf{v}\cdot\nabla\mathbf{v} = - \nabla q + a \mathbf{v} - b |\mathbf{v}|^2\mathbf{v} - (1 + \nabla^2)^2 \mathbf{v} \, , \qquad \nabla \cdot \mathbf{v} = 0 \, .
\end{equation}
The new coefficients are given by
\begin{equation}
\label{eq: coefficients rescaled one}
\lambda = k_\mathrm{c}\lambda_0/\sigma_\mathrm{c}\, , \qquad a = 1 - \alpha/\sigma_\mathrm{c}\, , \qquad b = \beta/\sigma_\mathrm{c}\ \, .
\end{equation}
In the following, we want to show how $\lambda$, $a$ and $b$ scale with the swimming speed $v_0$ by setting the other microscopic parameters to values appropriate for the bacterium \textit{Bacillus Subtilis}, based on recent experimental studies.
The length and diameter of \textit{Bacillus Subtilis} is approximately $\ell_0 = \SI{5}{\micro\meter}$ and $d = \SI{0.8}{\micro\meter}$~\cite{nishiguchi2018engineering}.
We set the density to $\rho = \SI{0.01}{\per\cubic\micro\meter}$, which has been used in experiments with bacterial suspensions in obstacle lattices~\cite{nishiguchi2018engineering}.
Further, we set $\epsilon = d = \SI{0.8}{\micro\meter}$, which corresponds to very short range interactions.
The three remaining parameters are tuned in such a way that the onset of pattern formation is found between $v_0 = \SI{10}{\micro\meter\per\second}$ and $\SI{20}{\micro\meter\per\second}$, which is consistent with experiments~\cite{sokolov2012physical}.
We use a viscosity ratio of $\mu_\mathrm{eff}/\mu_0 = 4$, a rotational relaxation time of $\tau = \SI{1}{\second}$ and an interaction strength of $\gamma_0 = \SI{1}{\per\micro\meter}$.
Note that these values are not supposed to be accurate, we rather use them to illustrate the dependencies of the macroscopic coefficients on the swimming speed $v_0$, which we vary from $10$ to $\SI{60}{\micro\meter\per\second}$, a range that is achievable by varying the oxygen concentration~\cite{sokolov2012physical}.
Fig.~\ref{fig: swimming speed three coeff} shows the coefficients $a$ and $b$ and nonlinear advection strength $\lambda$ as a function of the swimming speed $v_0$.
At the onset of the pattern formation, i.e., at the finite-wavelength instability~\cite{heidenreich2016hydrodynamic,reinken2019anisotropic,james2018turbulence}, $a$ becomes positive, which is shown in Fig.~\ref{fig: swimming speed three coeff}a) by the dashed line.
Initially, $a$ grows fast with the swimming speed, however, for larger values of $v_0$, the dependency is much less pronounced.
As Fig.~\ref{fig: swimming speed three coeff}b) shows, the coefficient $b$ is constant for the whole range of swimming speeds. 
The nonlinear advection strength, however, shows a linear increase with $v_0$, see Fig.~\ref{fig: swimming speed three coeff}c).
This clear dependency $\lambda(v_0)$ is consistent with an earlier derivation of a hydrodynamic model~\cite{bertin2009hydrodynamic} and suggests a way to test our predictions experimentally by using the swimming speed as control parameter.
By adjusting the oxygen concentration, one might be able to vary $v_0$ and investigate the transition from a disordered state to a stabilized vortex lattice.
We note that the values shown here are different from the values used for the numerical calculations. The above calculation rather serves as general argument for the dependencies on $v_0$.

\clearpage

\section{Impact of obstacle size}

\begin{figure}[h]
\includegraphics[width=0.5\linewidth]{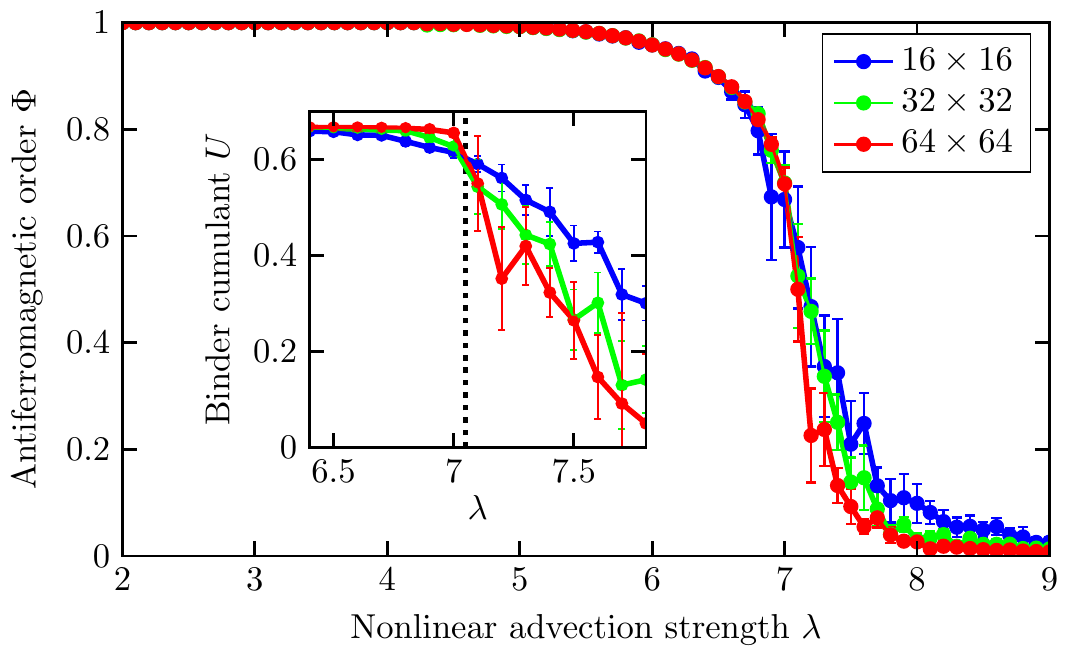}
\caption{\label{fig: order smaller pillars}Antiferromagnetic order parameter $\Phi$ as a function of nonlinear advection strength $\lambda$ for different system sizes $N=16\times 16$, $N=32\times 32$ and $N=64\times 64$ and an obstacle size of $\ell = 0.11 \Lambda$. We observe a transition from perfect antiferromagnetic order at low values of $\lambda$ to a disordered system at high values. The inset shows the Binder Cumulant as a function of $\lambda$. The intersection point of the curves for different system sizes marks the critical point $\lambda_\mathrm{c} \approx 7.1$, here denoted by the dashed line.}
\end{figure}

\begin{figure}[h]
\includegraphics[width=0.4\linewidth]{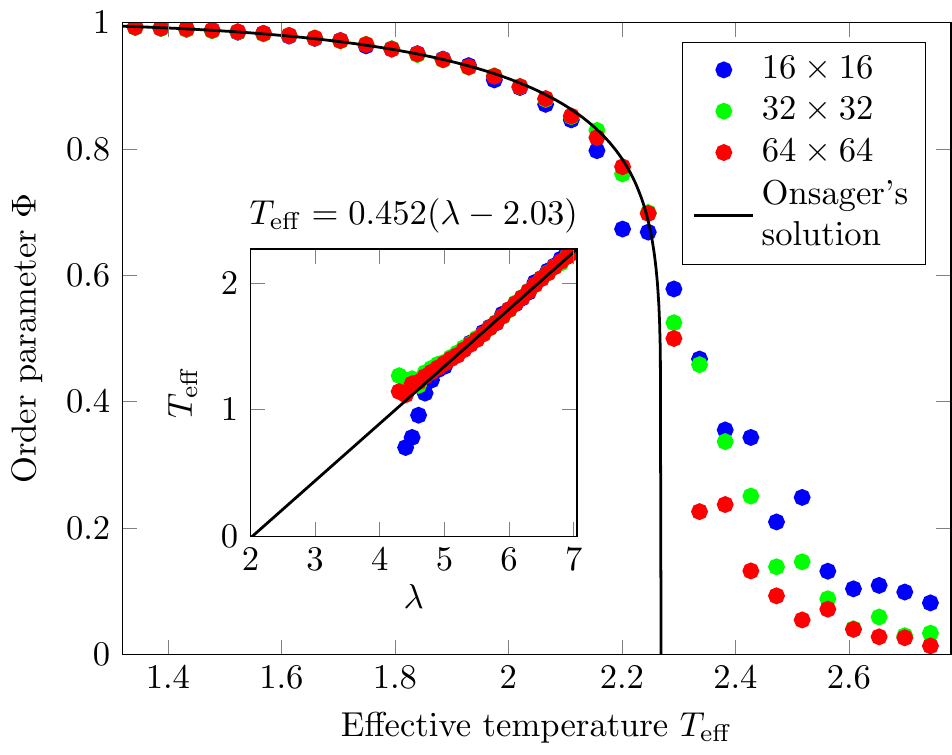}
\caption{\label{fig: effective temperature smaller pillars}Antiferromagnetic order parameter $\Phi$ as a function of effective temperature $T_\mathrm{eff}$ for different system sizes $N=16\times 16$, $N=32\times 32$ and $N=64\times 64$ calculated via comparison to Onsager's analytical solution (shown as solid black line). Here, the obstacle size is $\ell = 11 \Lambda$, in contrast to the Fig.~4 in the main text, where $\ell = 0.13 \Lambda$. The effective temperature $T_\mathrm{eff}$ is plotted as a function of $\lambda$ in the inset, where the solid black line shows the linear fit $T_\mathrm{eff} = 0.452(\lambda-2.03)$.}
\end{figure}

\begin{figure}[h]
\includegraphics[width=0.4\linewidth]{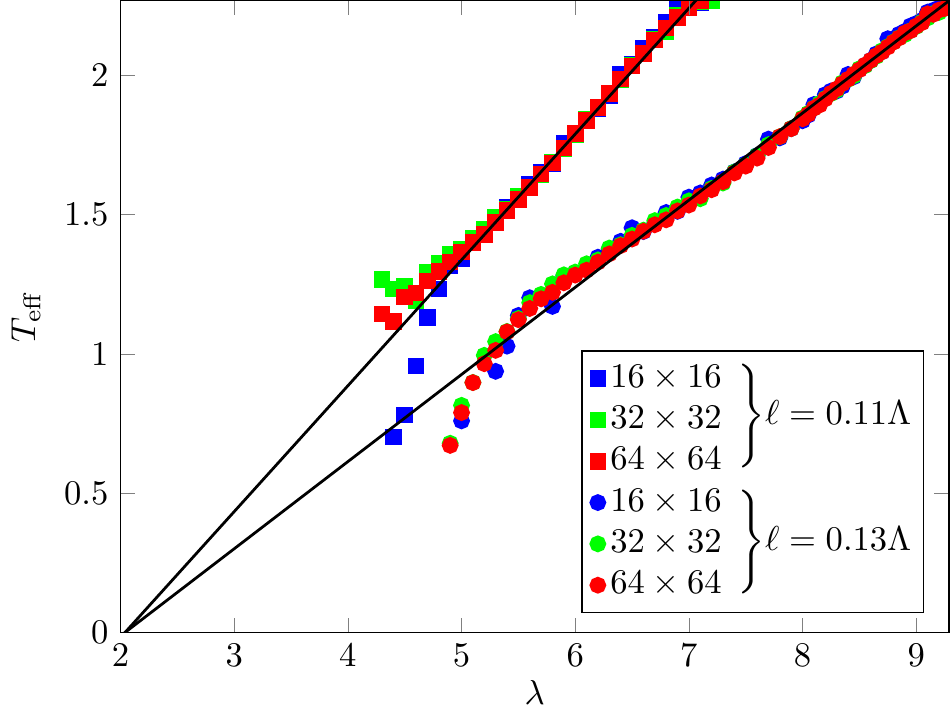}
\caption{\label{fig: comparison effective temperature pillar size}The effective temperature $T_\mathrm{eff}$ plotted as a function of $\lambda$ for different obstacle sizes $\ell = 0.11 \Lambda$ and $\ell = 0.13 \Lambda$ and system sizes $N=16\times 16$, $N=32\times 32$ and $N=64\times 64$. The solid black lines show the the linear fit. Decreasing the obstacle size shifts the critical point to smaller $\lambda$.}
\end{figure}

\begin{figure}[h]
\includegraphics[width=0.4\linewidth]{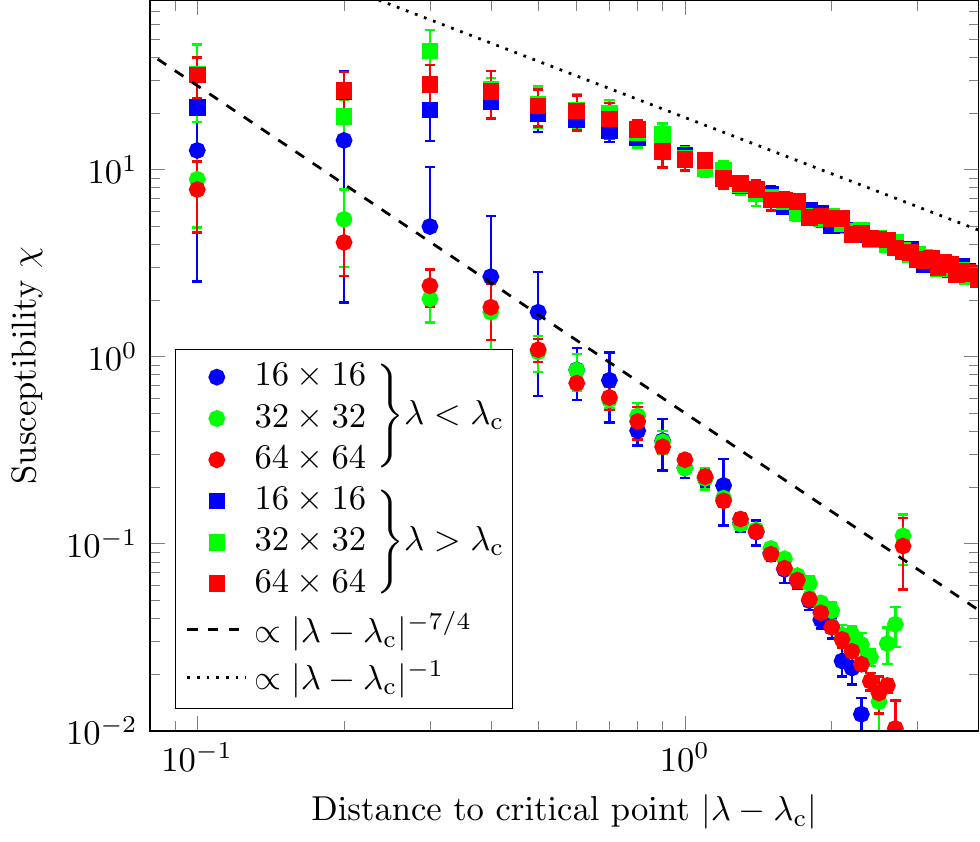}
\caption{\label{fig: scaling smaller pillars}Susceptibility $\chi$ as a function of the distance to the critical point $|\lambda-\lambda_\mathrm{c}|$ in a log-log plot for different system sizes $N=16\times 16$, $N=32\times 32$ and $N=64\times 64$ and an obstacle size of $\ell = 11 \Lambda$. Below the critical point, the susceptibility diverges with an exponent $\gamma\approx 7/4$ (slope shown as dashed line), consistent with the 2D Ising universality class. Above the critical point, however, we observe $\gamma = 1$ (slope shown as dotted line), which is consistent with the Curie-Weiss law. }
\end{figure}

For the results presented in the main text, we set the obstacle size to $\ell=0.13\Lambda$, which corresponds to the obstacle arrays investigated in recent experimental~\cite{nishiguchi2018engineering} and theoretical studies~\cite{reinken2020organizing}.
To determine the impact of obstacle size on the transition from a disordered to an antiferromagnetic vortex lattice, we repeat our analysis with a smaller obstacle size of $\ell=0.11\Lambda$, albeit with a smaller number of runs and therefore somewhat less reliable statistics.
Fig.~\ref{fig: order smaller pillars} shows the antiferromagnetic order parameter $\Phi$ and the Binder cumulant $U$ (in the inset) as a function of nonlinear advection strength $\lambda$.
Analogous to the approach outlined in the main text, we determine the critical point from the intersection point of the Binder cumulant curves for different system sizes and find $\lambda_\mathrm{c} \approx 7.1$.
This is significantly smaller than the value $\lambda_\mathrm{c} \approx 9.4$, which we obtained for the larger obstacles with $\ell=0.13 \Lambda$.
Again, comparing with Onsager's solution for the 2D Ising model with nearest-neighbor interactions (see Eq.~(5) in the main text), we calculate the effective temperature as a function of advection strength, yielding $T_\mathrm{eff}(\lambda) = 0.452(\lambda - 2.03)$.
This is shown in the inset of Fig.~\ref{fig: effective temperature smaller pillars}, where the accuracy of the linear fit is clearly visible.
Consistent with the case $\ell = 0.13 \Lambda$, absolute zero ($T_\mathrm{eff} = 0$) again corresponds to $\lambda^\star\approx 2$, which is where the unconstrained system starts to develop the turbulent state.
The slope of $T_\mathrm{eff}(\lambda)$, however, is increased significantly, which shows that a decrease in obstacle size shifts the transition to smaller values of $\lambda$.
This is probably due to the fact that the obstacles with $\ell = 0.11\Lambda$ have a smaller impact on the dynamics due their reduced size and, thus, are less efficient in stabilizing regular patterns.
To compare, we plot the dependence $T_\mathrm{eff}(\lambda)$ for both obstacle sizes $\ell=0.11 \Lambda$ and $\ell = 0.13 \Lambda$ in Fig.~\ref{fig: comparison effective temperature pillar size}.
Thus, slightly changing the properties of the obstacle lattice does not change the qualitative behavior, i.e., linear dependence $T_\mathrm{eff}(\lambda)$, but changes the quantitative mapping $\lambda \rightarrow T_\mathrm{eff}$.
Further, Fig.~\ref{fig: scaling smaller pillars} shows the susceptibility $\langle \chi \rangle$ as a function of the distance to the critical point $|\lambda - \lambda_\mathrm{c}|$.
We observe the same qualitative behavior as for $\ell=0.13\Lambda$, i.e., a scaling exponent consistent with the 2D Ising universality class ($\gamma \approx 7/4$) below the critical point and consistency with the Curie-Weiss law ($\gamma=1$) far above the critical point in the disordered state.

\clearpage

\section{Impact of lattice constant}

\begin{figure}[h]
\includegraphics[width=0.5\linewidth]{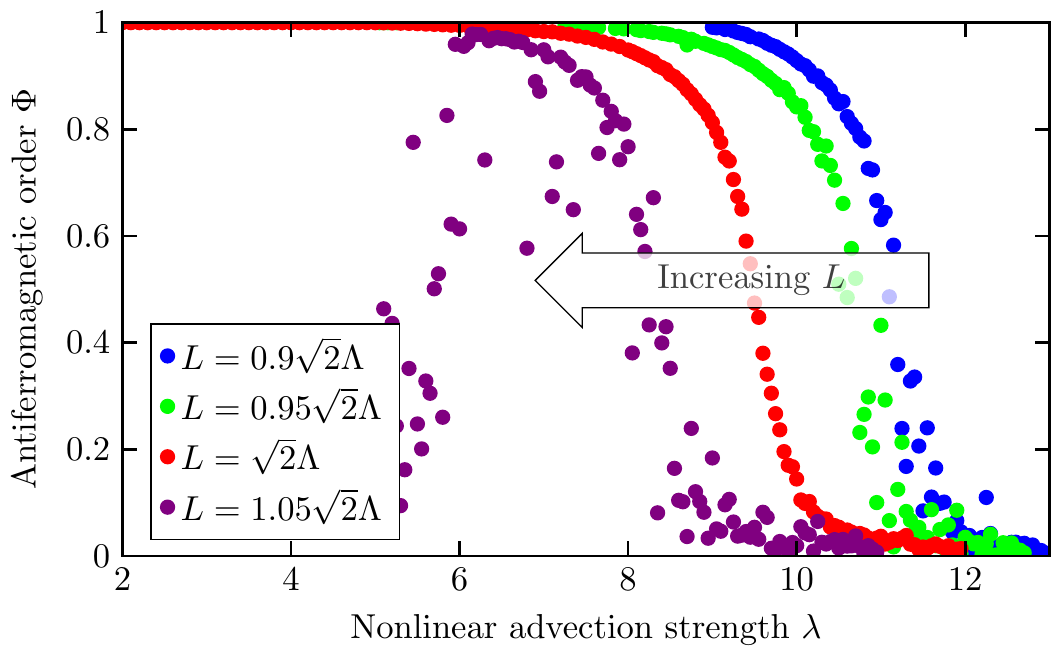}
\caption{\label{fig: compare lattice constant}Time-averaged antiferromagnetic order parameter $\Phi$ as a function of nonlinear advection strength $\lambda$ for different lattice constants $L = 0.9 \sqrt{2}\Lambda$, $L=0.95\sqrt{2}\Lambda$, $\Lambda = \sqrt{2}\Lambda$ and $\Lambda = 1.05\sqrt{2}\Lambda$. The system size is $N=32\times 32$ and the obstacle size is $\ell = 13 \Lambda$. Decreasing the lattice constant from $L=\sqrt{2}\Lambda$ (which fits exactly to the fastest growing mode) to slightly smaller values shifts the transition to larger $\lambda$, whereas an increase in $L$ shifts the transition to smaller $\lambda$.}
\end{figure}

For the results presented in the main text, we set the lattice constant to $L=\sqrt{2}\Lambda$, which corresponds to the case where the distance between obstacles fits exactly to the wavelength of the fastest growing mode in the unconstrained system~\cite{reinken2020organizing}.
Fig.~\ref{fig: compare lattice constant} shows the impact of changing the lattice constant slightly.
Increasing the lattice constant shifts the transition to smaller values of nonlinear advection strength $\lambda$, i.e., the stabilization of an antiferromagnetic vortex lattice is not as effective as for $L=\sqrt{2}\Lambda$.
This is expected for two reasons.
First, the characteristic length scale $\Lambda$ of the emerging patterns does not fit perfectly to the distance between obstacles, i.e., the emerging vortices are smaller than the gaps.
Second, the stabilizing effect of the obstacle lattice becomes smaller as the density of obstacles decreases.
The disorder at lower $\lambda$ for $L=1.05\sqrt{2}\Lambda$ is due to the slight misfit between $\Lambda$ and $L$, which leads to domains with different phases yielding zero global magnetization.
In contrast, when the lattice constant is decreased, the transition shifts to higher values of $\lambda$, i.e., the stabilization of an antiferromagnetic vortex lattice is even more effective.
This is probably due to the higher density of obstacles which increases the stabilizing effect. 
Decreasing the lattice constant even further, beyond $L < 0.9\sqrt{2}\Lambda$, the obstacles are so close together that the patterns are unable to grow and we do not observe vortices emerging in the gaps.
For a more detailed discussion on the impact of the lattice constant on the emerging vortex patterns in bacterial suspensions in obstacle arrays, see~\cite{nishiguchi2018engineering,reinken2020organizing}.

\clearpage

\section{Lattice with obstacle vacancies}

\begin{figure}[h]
\includegraphics[width=0.38\linewidth]{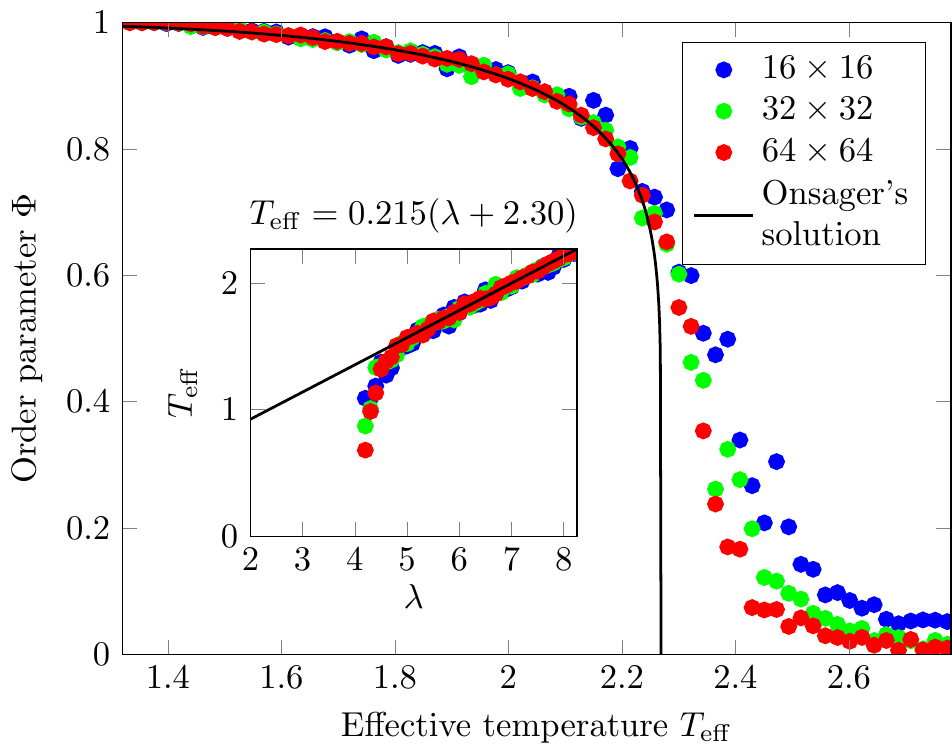}
\caption{\label{fig: effective temperature diluted}Antiferromagnetic order parameter $\Phi$ as a function of effective temperature $T_\mathrm{eff}$ for different system sizes $N=16\times 16$, $N=32\times 32$ and $N=64\times 64$ calculated via comparison to Onsager's analytical solution (shown as solid black line) in an obstacle lattice with $\SI{5}{\percent}$ vacancies. The effective temperature $T_\mathrm{eff}$ is plotted as a function of $\lambda$ in the inset, where the solid black line shows a linear fit.}
\end{figure}

\begin{figure}
\includegraphics[width=0.38\linewidth]{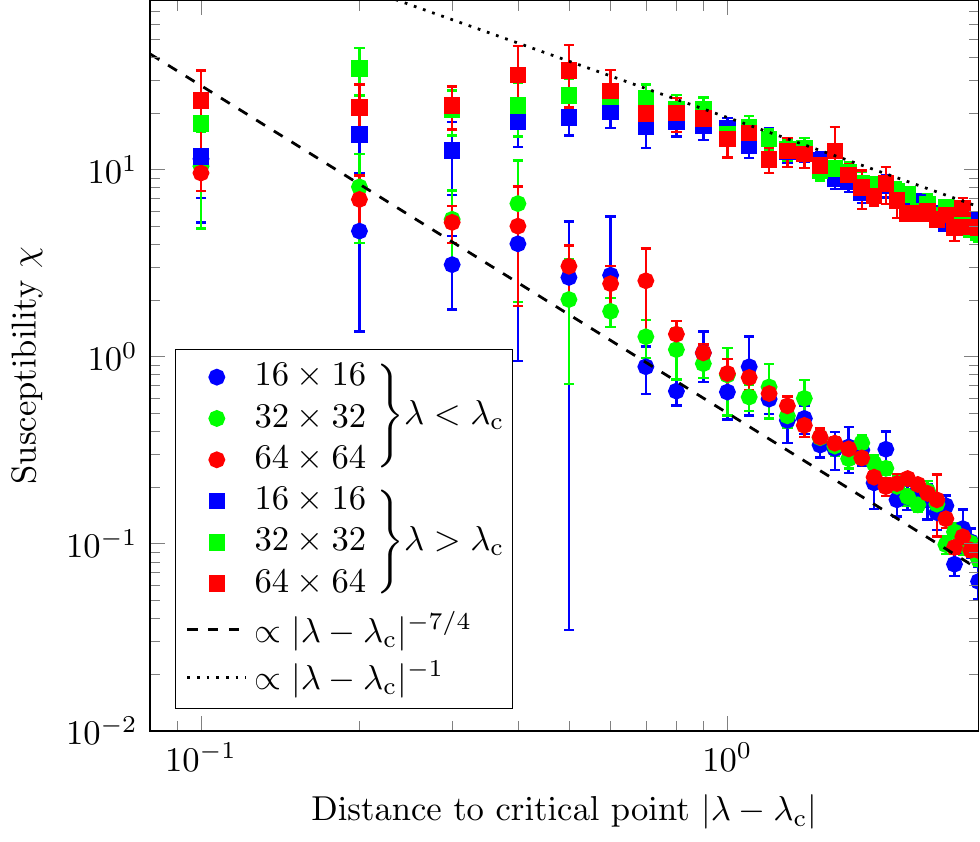}
\caption{\label{fig: susceptibility diluted}Susceptibility $\chi$ as a function of the distance to the critical point $|\lambda-\lambda_\mathrm{c}|$ in a log-log plot for different system sizes $N=16\times 16$, $N=32\times 32$ and $N=64\times 64$ in an obstacle lattice with $\SI{5}{\percent}$ vacancies. Below the critical point, the susceptibility diverges with an exponent $\gamma\approx 7/4$ (slope shown as dashed line), consistent with the 2D Ising universality class. Above the critical point, however, we observe $\gamma = 1$ (slope shown as dotted line), which is consistent with the Curie-Weiss law. }
\end{figure}

\begin{figure}
\includegraphics[width=0.38\linewidth]{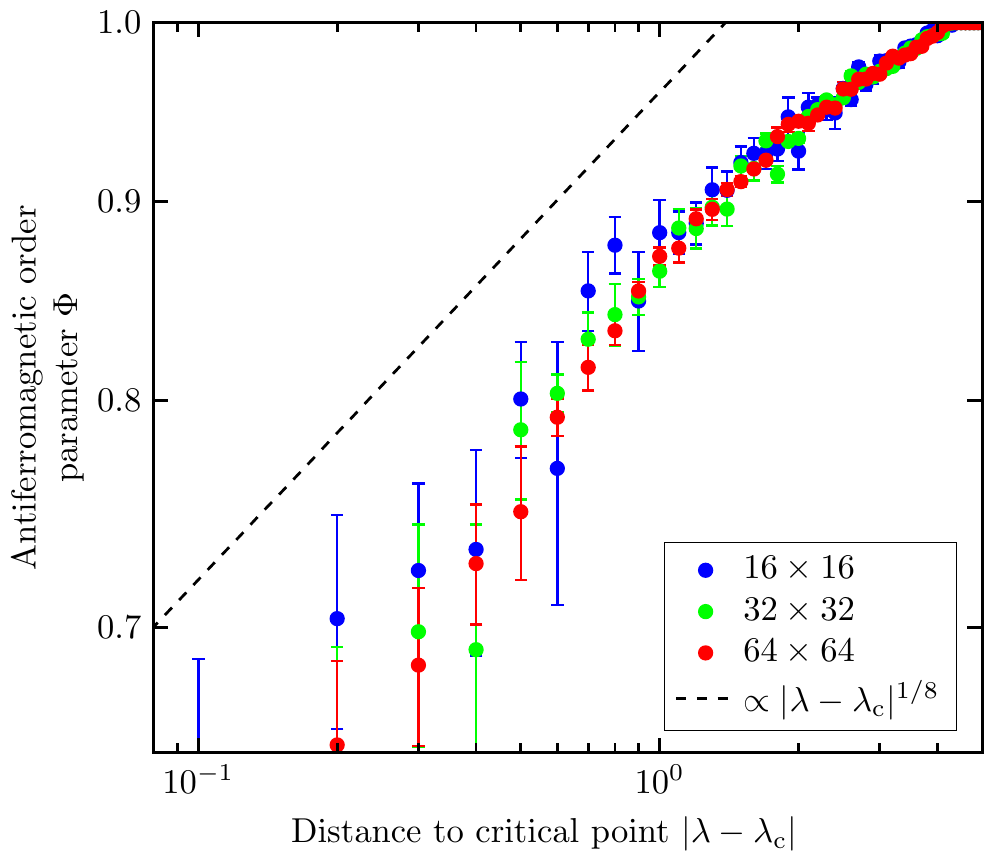}
\caption{\label{fig: order log log diluted}Antiferromagnetic order parameter $\Phi$  as a function of $|\lambda - \lambda_\mathrm{c}|$ for different system sizes $N=16\times 16$, $N=32\times 32$ and $N=64\times 64$ in an obstacle lattice with $\SI{5}{\percent}$ vacancies. Power-law behavior with exponent $\beta = 1/8$ is shown as a dashed line. The error bars represent the standard error.}
\end{figure}

In order to further test the robustness of our results, we further performed numerical calculations for a system where the obstacle lattice has defects, i.e. a small number of obstacle vacancies.
To this end, we randomly remove $\SI{5}{\percent}$ of the obstacles of the lattice.
We use the same lattice constant $L=\sqrt{2}\Lambda$ and the same obstacle diameter $\ell = 0.13\Lambda$ as for the calculations presented in the main text.
Introducing a small number of obstacle vacancies shifts the critical point to smaller values of nonlinear advection strength $\lambda$, but does not change the nature of the transition.
Calculating the Binder cumulant, we identify the critical $\lambda$ as $\lambda_\mathrm{c} \approx 8.5$, compared to $\lambda_\mathrm{c}\approx 9.5$ in the lattice without vacancies.
Thus, the incomplete obstacle lattice is less efficient in stabilizing the ordered state.
Using the same approach as before, we again find a linear relation between effective temperature and nonlinear advection strength, specifically $T_\mathrm{eff} = 0.215(\lambda + 2.30)$.
Fig.~\ref{fig: effective temperature diluted} shows the antiferromagnetic order parameter $\Phi$ as a function of $T_\mathrm{eff}$.
The relation $T_\mathrm{eff} (\lambda)$ is shown in the inset.
The consistency with the Ising universality class is further illustrated in Figs.~\ref{fig: susceptibility diluted} and \ref{fig: order log log diluted}, where the susceptibility $\chi$ and the order parameter $\Phi$ are plotted as a function of the distance to the critical point $|\lambda - \lambda_\mathrm{c}|$, respectively.
Both quantities show the same scaling behavior as for the lattice without vacancies.
We observe a scaling exponent of the susceptibility $\chi$ consistent with the 2D Ising universality class ($\gamma \approx 7/4$) below the critical point and consistency with the Curie-Weiss law ($\gamma=1$) far above the critical point in the disordered state.
The antiferromagnetic order parameter $\Phi$ scales with $\beta = 1/8$ close to the transition, indicating Ising universality.
This analysis shows the robustness of the results with respect to the addition of a small amount of defects into the system.
Similar kinds of disorder have been introduced to the Ising model, where it has been found that the universality classification is independent of the dilution for bond- and site-diluted systems~\cite{martins2007universality,kenna2008scaling}.
The addition of a random field, however, leads to a more complex behavior of the Ising model in two dimensions characterized by the absence of a globally ordered state~\cite{grinstein1984lower}.
Removing obstacles does not directly correspond to any of the above-mentioned cases as we do not influence the spins or the interactions directly but rather implicitly by reducing the constraints on the flow field.
However, our results indicate that adding obstacle vacancies is somewhat similar to the bond- or site-diluted Ising model, at least for the small amount of disorder introduced here.

\clearpage
\input{supplemental.bbl}

%% file: main.bbl
%

%% file: supplemental.bbl
%